\documentclass[acmtog,preprint]{acmart}
\acmSubmissionID{404}
\usepackage{adjustbox}
\usepackage{booktabs} 

\setcitestyle{nosort,square} 

\def\code#1{{\relsize{-1}{\texttt{#1}}}}

\usepackage{relsize}
\usepackage[ruled]{algorithm2e} 
\renewcommand{\algorithmcfname}{ALGORITHM}
\SetAlFnt{\small}
\SetAlCapFnt{\small}
\SetAlCapNameFnt{\small}
\SetAlCapHSkip{0pt}

\copyrightyear{2022}
\acmYear{2022}
\setcopyright{acmcopyright}
\setcopyright{none}
\acmConference{Conference Name}{Conference Date and Year}{Conference Location}
\acmDOI{10.1145/8888888.7777777}
\acmISBN{978-1-4503-1234-5/22/07}

\def\island{\emph{island}\xspace}

\def\landscape{\emph{landscape}\xspace}

\definecolor{OliveGreen}{RGB}{40,120,20}

\begin{document}
\makeatletter
\title{Data Parallel Path Tracing in Object Space}
\author{Ingo Wald}
\affiliation{NVIDIA}
\author{Steven G Parker}
\affiliation{NVIDIA}

\begin{teaserfigure}
\centering
\setlength{\tabcolsep}{1mm}
\begin{tabular}{cc}
\includegraphics[width=.49\textwidth]{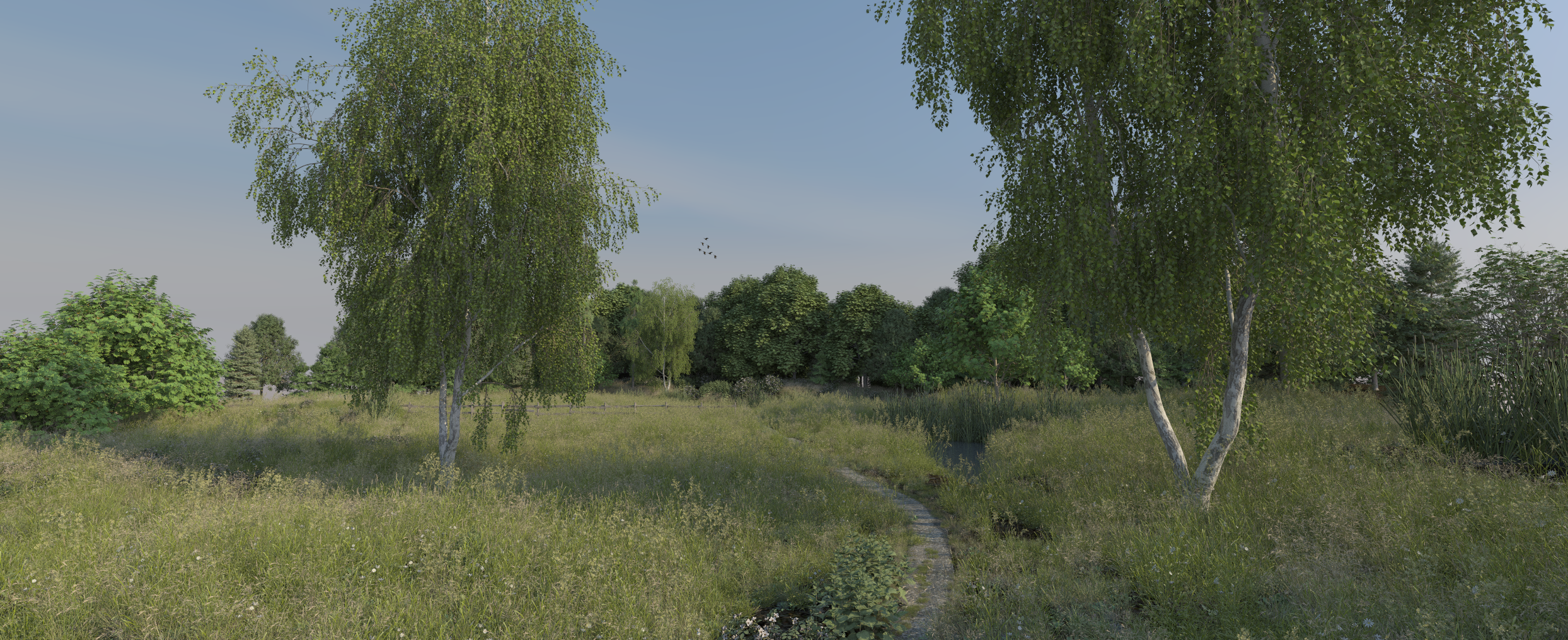} &
\includegraphics[width=.49\textwidth]{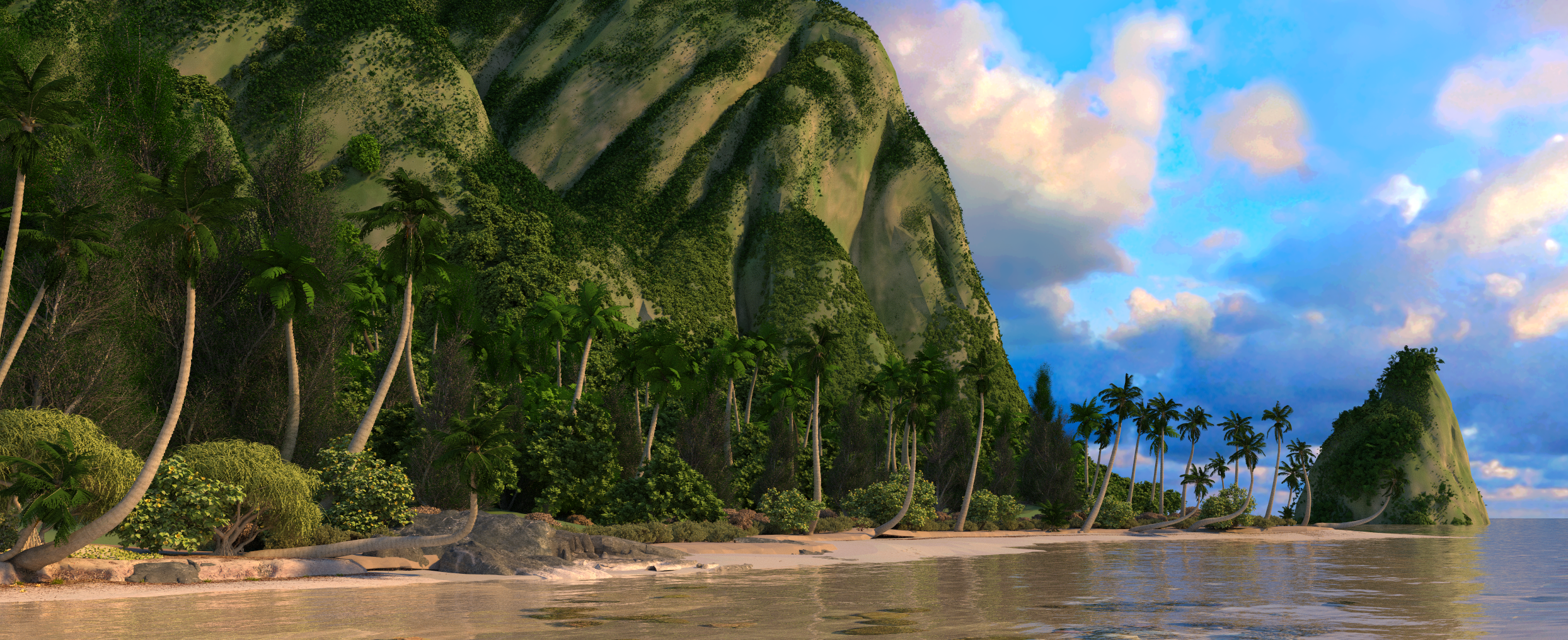}   \\
\hline
\multicolumn{2}{c}{(Hardware: 4 worker nodes w/ $2\times$RTX8000, low-end head node, 10-Gigabit Ethernet, screen size $2560\times 1080$)}
\\
\hline
PBRT \emph{landscape}: max 3.7~GB GPU usage, 6.2~FPS (1 path/pixel)
&
Disney Moana \emph{island}: max 25~GB GPU usage, 7.9~FPS (1 path/pixel)
\\
\relsize{-1}{30~K instances, 4.3~B inst.~triangles, 370 unique meshes, 500~MB image textures} &
\relsize{-1}{39~M instances, 41~B inst.~triangles, 7~M unique meshes, 804~MB baked-PTex txt.} \\
\end{tabular}
\vspace{-1em}
\caption{\label{fig:teaser} Two screenshots from a 
  data-parallel path tracer built using the techniques described in
  this paper; showing multi-bounce path tracing, textures, alpha
  textures, area- and environment lighting, etc., on two
  non-trivial models each distributed across 4 nodes and 8 GPUs.  Despite
  intentionally low-end network infrastructure, at $2560\times 1080$ pixels and one path per pixel
  these two examples run at 6.2 and 7.9 frames per second, respectively
  (images shown are converged over multiple frames).  }
\end{teaserfigure}

\begin{abstract}
We investigate the concept of rendering production-style content with
full path tracing in a data-distributed fashion---that is, with
multiple collaborating nodes and/or GPUs that each store only part of the
model. In particular, we propose a new approach to tracing rays across
different nodes/GPUs that improves over traditional spatial partitioning,
can  support both object-space and spatial partitioning (or any
combination thereof), and that enables multiple techniques
for reducing the number of rays sent across the
network.  We show that this approach can handle different kinds of
model partitioning strategies, and can ultimately render non-trivial
models with full path tracing even on quite moderate hardware
resources with rather low-end interconnect.
\end{abstract}


\maketitle

\section{Introduction}

Data-parallel (or data-\emph{distributed}) rendering is the process of
rendering a model whose constituent components are distributed across
the memories of multiple different compute units such as HPC compute
nodes or GPUs (which, following MPI parlance, we will often
call \emph{ranks}). This is typically done for one of two  reasons:
one is that a given model is too large to fit into the memory of a
single node, and the user \emph{chooses} to distribute it across
several different nodes (we call this \emph{explicit} distribution);
the second is that for some reason or other the model's data is
already spread across different nodes, and cannot easily be
merged for rendering (i.e., it is \emph{natively} distributed).
Data-parallel rendering did attract some attention in the past, but in
practice today it is almost entirely confined to scientific
visualization (\emph{sci-vis}), and even there is almost only
performed with image compositing-based approaches that cannot handle
effects like path tracing. For the kind of images shown in
Figure~\ref{fig:teaser}, data parallel rendering today is virtually
non-existent. The list of possible explanations for that is long, and
a full discussion beyond the scope of this paper. However, we argue
that these reasons can be sorted into either one of two groups: one
that argues that there is neither need nor demand for data-parallel
rendering in production rendering; and one that argues that it is too
hard.

For the first one, we argue that data is increasingly moving into
the cloud (where parallel resources are easily available), and that
content is continuing to grow at a rate that far surpasses the rate
at which GPU or even host memories are growing. The second is
more interesting, as the data parallel rendering techniques we use
in sci-vis today  may indeed not be ideally suited for this context:
First, sci-vis rendering largely relies on
image compositing, but for path tracing we certainly can not. Path
tracing requires the frequent forwarding of either rays or data between nodes; and that is expensive.
Second, the content used in production
rendering is very different than that encountered in visualization, 
including spatially large yet hard to split instances or meshes with shading data,
abundant spatial overlap, 
etc. 
As we show below this content does not always do well with
spatial subdivision, yet this is what virtually all data-parallel rendering
today is built on.

In this paper, we take a closer look at data-parallel
path tracing for production style content. In particular, we 
 borrow some of the last few years' insights on
ray tracing acceleration structures, and use that to
propose several new techniques 
that go beyond strictly spatial scene subdivisions. We do this through a
combination of two things: First, we describe distributed content
through what we call \emph{proxies}--bounding boxes that describe which
parts of the model can be found on which rank(s), and which are allowed to
arbitrarily overlap any other proxies.
Second, we propose an object-space
parallel \emph{ray forwarding operator} that---using those proxies---enables
each ray to easily determine which node/GPU it should next be forwarded to.
We describe several techniques that make this efficient, and demonstrate this
using a prototype data parallel path tracer built using these techniques (Figure~\ref{fig:teaser}).

\ifx\empty
\subsection*{Contributions}
 
 \begin{itemize}
 \item An argument for thinking beyond purely spatial subdivision for
   data-parallel rendering; and a set of techniques to help realize
   this new paradigm
 \item A new, general formulation of how model data is owned across
   different ranks using proxies---a set of bounding boxes with
   ownership information that can compactly yet tightly represent
   distributed data; proxies can overlap other proxies from both same
   and other nodes, and can represent both spatial and object space
   partitions as well as any combination thereof.
 \item The introduction of the concept of a \emph{local ray-forwarding
   operator} that determines where a given ray needs to go to next,
   and one realization of this operator that minimizes the number of
   ray forwards.
 \item A description of how these concepts can be used to realize a
   complete data-parallel GPU path tracer for realistic content.
 \item A proof of concept evaluation of this framework that
   demonstrates the potential of this approach.
 \end{itemize}
\fi

\section{Related Work}

Parallel rendering refers to a family of methods where multiple nodes
and/or GPUs work together to render one model; \emph{data-parallel}
rendering to where every node has only part of the model.  Throughout
this paper we adopt the parlance of the Message Passing Interface
(MPI)~\cite{MPI,MPI2}, referring to \emph{ranks} in \emph{groups} that
can communicate with each other through the passing of messages. MPI
can also run multiple ranks on the same node, but we will typically
use one rank per GPU.

\begin{figure*}[h]
\includegraphics[width=.246\textwidth]{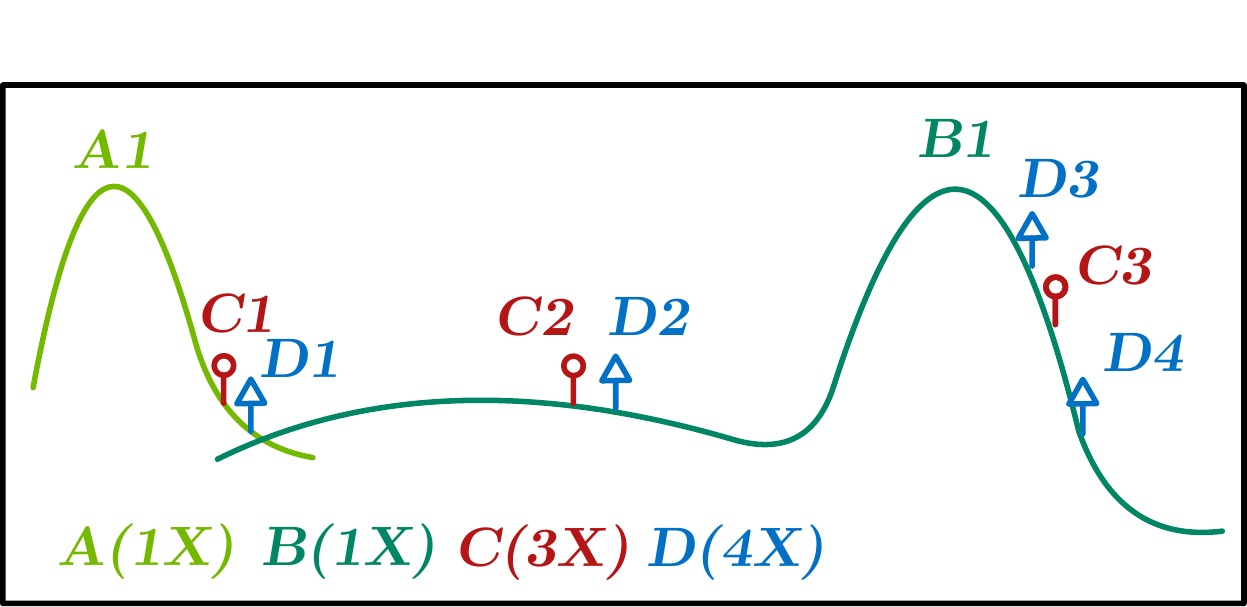}
\includegraphics[width=.246\textwidth]{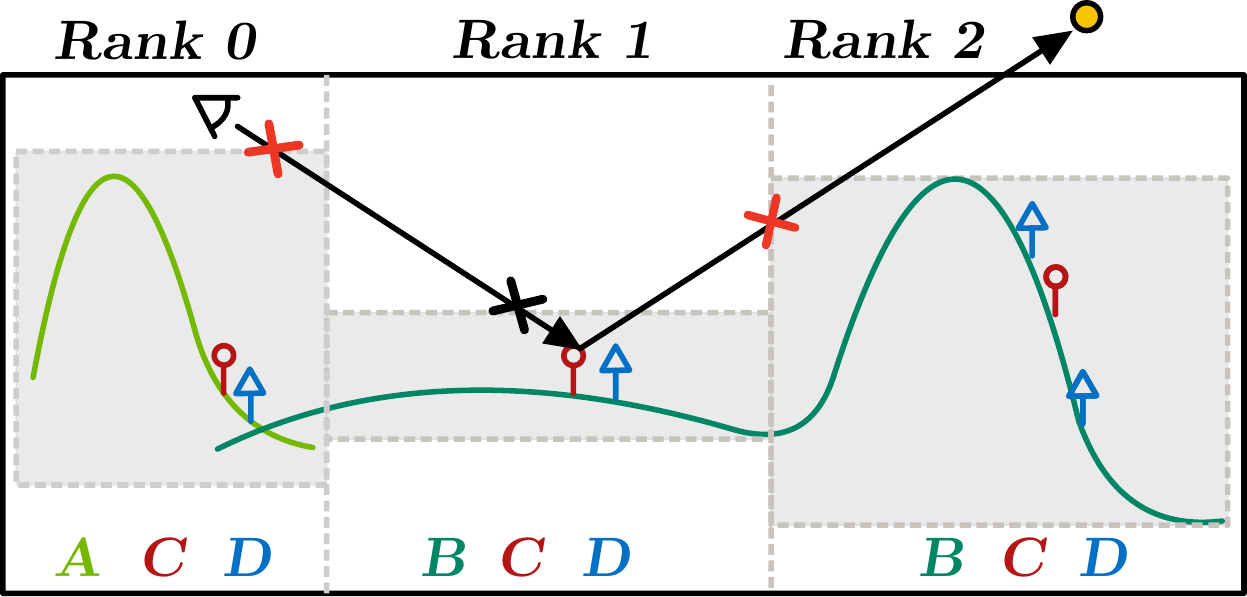}
\includegraphics[width=.246\textwidth]{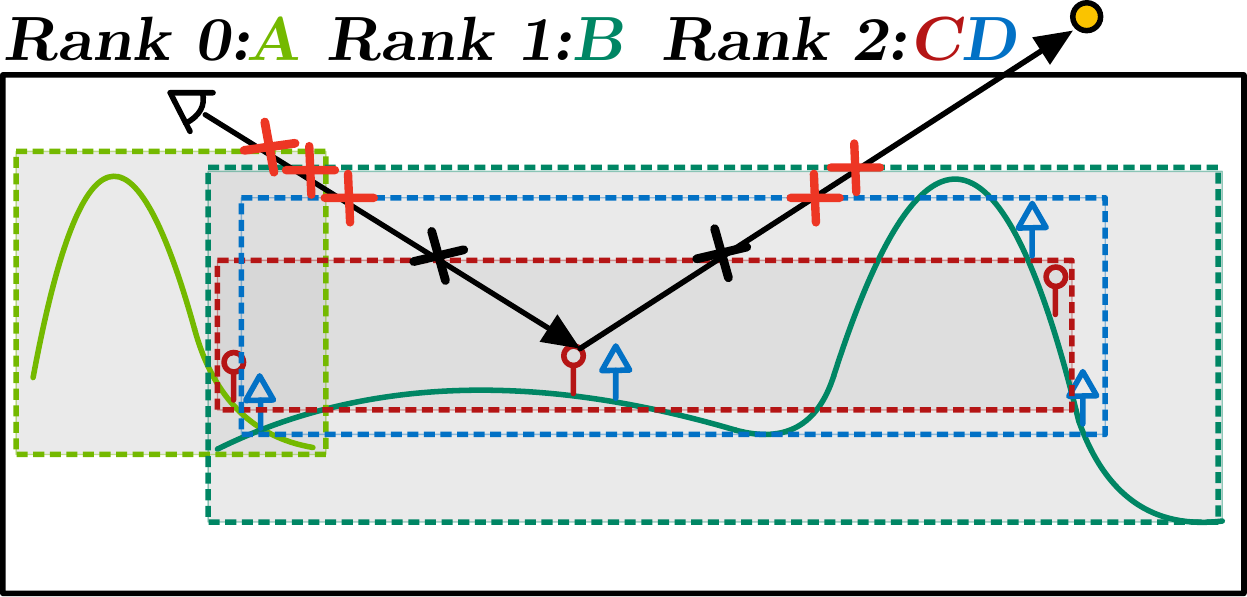}
\includegraphics[width=.246\textwidth]{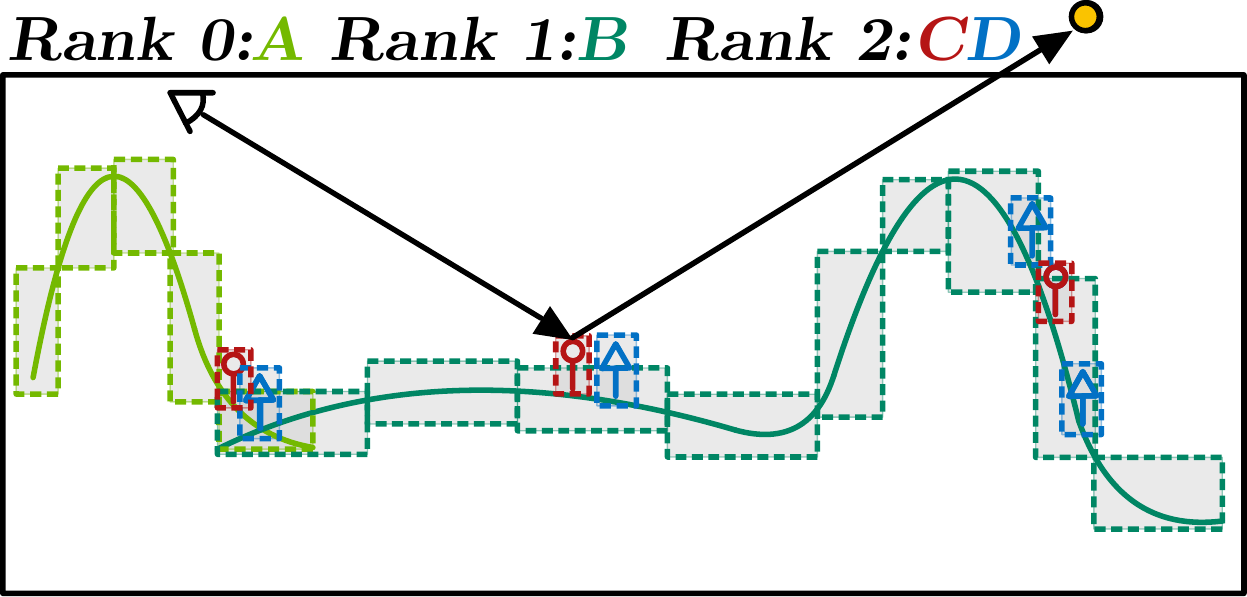}
\\[-1em]
\caption{\label{fig:sketch} 
\ifx\empty
Sketch of the core ideas behind our
  method, illustrated on a simplified model that is conceptually
  similar to the \island model (a). The illustrated model consists of
  two base meshes (A and B), three instances of tree C and four of tree D.  
  b) spatial partitioning has two issues with this
  kind of data: it struggles to properly partition it with spatial
  splits, and it can end up with large boxes that require lots of ray
  forwarding between ranks. c)~Switching to object space partitioning
  where different partitions allow for doing non-spatial splits, but
  if done naively ends up with even worse ray traffic. d)~we represent
  objects by more than one---possibly overlapping, but
  tight-fitting---''proxies'': these are both more flexible and
  general than either spatial or object space techniques alone, while
  simultaneously being tighter and producing less ray traffic.
  \vspace{-1em}
  \else
  The core ideas behind our
  method, illustrated on a 2D sketch of a model very like the \island model. a) 
  Our model consists of
  two base meshes (A and B), and three respectively four instances of two types of tree (C and D).  
  b) spatial partitioning on this model runs into two issues: it struggles to properly partition it with spatial
  splits, and it can end up with large boxes that 
  incur ray
  forwarding between ranks. c)~Object space partitioning
  allow for non-spatial scene partitions, but
  if done naively ends up with even worse ray traffic. d)~we represent
  objects by more than one---possibly overlapping, but
  tight-fitting---''proxies'': these are more flexible and
  general than either spatial or object space techniques, and also require less forwarding.
  \vspace{-1.8em}
  \fi
}
  
\end{figure*}

Today, data parallel rendering is
almost entirely confined to svi-vis, with packages like
ParaView~\cite{paraview} and VisIt~\cite{visit} handling the data
distribution, and communication-optimized libraries such as
Ice-T~\cite{ice-t} handling the compositing.  This is used for both
volumetric and polygonal data, but is generally restricted to simple
shading where alpha- or Z-compositing can be used.
Sci-vis also uses path tracing, but
in this case usually relies on data replicated rendering.

In a ray tracing context, parallel rendering is easiest to realize
when using either replicated rendering~\cite{wald-moana,facebook-cluster}.
Approaches to
dealing with models larger than available
memory can be classified into three categories: \emph{out of
  core} ray tracing, where data is paged in on demand, usually
including some form of batching, sorting, scheduling, and
caching~\cite{pharr-mcrt,hyperion}; \emph{data forwarding}, where data
is sent to / fetched by whatever node needs
it~\cite{wald-caching,demarle-caching,ize-caching,dgx-pt}; and 
 \emph{ray forwarding} where rays get scheduled on, and sent to,
other node(s)~\cite{hypercube-rt,reinhard::PhD,kilauea,spray,galaxy}. 

An interesting exception to this classification is the concept
of hardware-assisted \emph{distributed shared memory}
as recently used by Jaros et al.'s~\shortcite{dgx-pt}, in their case using 
CUDA unified memory and NVLink on an NVidia DGX: on the hardware
level this uses data-forwarding over high-bandwidth NVLink, with unified memory
giving the appearance of a single replicated address space; on the software side,
the authors show that best performance is achieved if the application is aware of which
data lives on which physical GPU, and ideally even replicates some of that data.

Looking at the literature on data parallel ray tracing we make two observation:
First, that the ray forwarding approach has received but scant
attention: some early work has looked
at the scheduling part of the problem (see, e.g.,~\cite{reinhard::PhD}
for an overview), but we are aware of only three somewhat-recent
approaches that actually forward rays: The \emph{Kilauea}
renderer~\cite{kilauea} (which simply forwarded every ray to every
node), Park's \emph{SpRay} system~\shortcite{spray} (which focuses
primarily on scheduling and speculative execution), and TACC's
\emph{Galaxy}~\cite{galaxy} (which uses techniques similar to those in
SpRay). Second, across all the different approaches to data parallel
rendering taken over the last four decades researchers seem to have
taken it for granted that data would necessarily get \emph{spatially}
partitioned using various forms of grids, octrees, kd-trees, etc. The
latter is particularly striking given the last decade's lively
discussion around spatial vs object hierarchies in ray tracing. In
this field, the community has largely switched from spatial techniques
to object-space techniques like
BVHes~\cite{optix,embree,turing}. These BVHes are often built using
spatially influenced techniques like top-down partitioning and the
Surface Area Heuristic (SAH)~\cite{sah-bvh}), and do best when
augmented with optimizations like spatial
splits~\cite{spatial-splits-1,spatial-splits-2,spatial-splits-3} and
braiding~\cite{braiding}---but they are nevertheless object-space
techniques.

Issues with spatial partitioning have also been
pointed out by Zellmann et al.~\shortcite{zellmann}. They
proposed techniques that produce better spatial partitions, but never
went beyond spatial partitions.

\section{Data Parallel Ray Traversal
}

In this section, we will give a high-level overview of the general
concepts and techniques for what we call \emph{data-parallel ray
  traversal} in object space. The core idea is to combine two things:
first, a more general representation of data-distributed scene content
where different pieces of content across different nodes are allowed to
spatially overlap other such content, both on the same and/or other
nodes; or even to be replicated across multiple nodes if and where
desired. Second, the concept of what we call a \emph{data-distributed
  forwarding operator} that, in the presence of such object-space
distributed content can always tell which other node a given ray needs
to be forwarded to next, in a guaranteed correct yet efficient manner
that explicitly aims to minimize the number of times that such
forwarding needs to happen.

\subsection{Goals, Non-Goals, and Key Issues}

Our goal is to render---with full path tracing---models that are larger
than the memory of any one of our ranks (and in general, larger than the host memory of the nodes that contain these GPUs).
We explicitly target assets similar to the \island model
that---at least with a good partitioner---may today need only on the
order of four or eight GPUs; not the ``at scale'' type visualizations
of up to thousands of nodes that are so common in visualization. While
we do believe our method will also work in larger scenarios these
are not currently our focus; nor is the concept of strong vs weak
scaling that is so important in sci-vis.

We also observe that such content is very
 different that
 encountered in sci-vis.  For
example, spatial overlap of different logical objects and instances is
not just a possibility, but a given. 

While we do aim for interactive performance, we do not (yet) aim for fully real-time photo-realistic
rendering. We do believe our method to be a first step towards this
goal, but this will ultimately require more systems work than
entertained in this paper.
We expect network bandwidth to be the ultimate bottleneck, but
we still need to perform a fair amount of shading and
texturing. Thus, though our framework can also be recompiled to a
CPU-only backend using Embree for this paper we
only consider  GPUs.

With the growing gap between compute and
bandwidth our main concern is to
reduce the amount of network bandwidth required for a given frame.
This can lead to un-intuitive situations: in data replicated rendering one can expect  that adding
more resources will improve performance, but in data-parallel
the opposite is often the case---adding more ranks 
also increases the chance that rays need to get forwarded, which is
counter-productive. 

\subsection{Core Idea}
\label{sec:idea}

With these goals in mind, let us take a look at how state of the art
data-parallel ray tracing would work for the kind of content we are
targeting: Let us consider a simplified example of a 2D "island" model
shown in Figure~\ref{fig:sketch}a: a model made of two base meshes,
with two types of trees that each have several instances.

Let us now first consider the state of the art, and let us create an
imaginary spatial partitioning of this model into three disjoint
regions (Figure~\ref{fig:sketch}b): no matter where exactly the splits
are placed in this example, almost every spatial domain is overlapped by almost every base
mesh, and contains at least one copy of each type of tree. Though the
number of \emph{instances} in each region has decreased,
memory per node barely has.
Let us now also look at two 
hypothetical rays as shown in Figure~\ref{fig:sketch}b: traversing
these rays
across those spatial domains is obvious and trivial, but each of these
rays touches two ranks, despite not even being close to any of the
geometry in the first or last rank (this only gets worse in 3D).

Let us now consider that same model with a
purely object-space partitioning, and simply assign the first mesh to
one rank, the second to another, and all instances of all trees to the
third (Figure~\ref{fig:sketch}c). Replication is now gone completely,
but overlapping boxes mean that rays are now touching even more
ranks' data; and traversal order is no longer obvious.  Now
let us take this general idea, but apply some ideas from the last
decade's discussion on BVHes vs kd-trees; in particular, concepts such
as \emph{spatial splits}~\cite{spatial-splits-1,spatial-splits-2} and
\emph{braiding}~\cite{braiding} to reach ``into'' large instances, and
instead represent them with several individual, tight-fitting bounding
boxes. If we do this (Figure~\ref{fig:sketch}d) an indeal algorithm
could now---if it existed---trace both of these rays on the same rank, without
any ray forwards at all.
\vspace{-1ex}

\subsection{Proxies and Data-Distributed Traversal}

Our initial plan to realize the ideas sketched in the previous section
was to assign each instance to exactly one rank, to create exactly one
proxy per instance, and to have each rank build exactly the same BVH
over those proxies. This suggests an obvious BVH-style traveral through
those proxies, sending rays to the nodes that owned the proxies they
traversed.  To have the next node continue traversal where the
previous one left off we had planned on using a stack-free BVH
traversal such as described by Hapala~\shortcite{wald-stackless} or
Vaidyanathan~\cite{short-stack}. This is indeed a useful mental
picture of how our method works; however, we can significantly improve
upon this as described in the rest of this section.

\subsubsection{Traversing nodes, not instances.}
Once a ray hitting a given proxy is sent to the node owning that
proxy there is no reason to limit intersection to that one
instance that generated that proxy. Local geometry intersection is
cheap compared to sending a ray over the network, so we should
always intersect all geometry on that node.  Once we do that we want
to make sure that no matter which proxies a ray encounters it will
never be sent to the same node twice.  This requires tracking the nodes
a ray has already visited; we present two options for that
below.

\subsubsection{Front to back traversal.}
A stack-free BVH-style traversal of the proxies sounds easy, but in a
distributed context it isn't, as even minute differenced in different
ranks' BVHes could lead to infinite loops; nor would it guarantee
front-to-back traversal. If, however, rays do keep track of which
nodes they have already visited, then we can do something that is even
simpler: we simply trace the ray into those proxies, and find the
closest such proxy that belongs to a node that we have not yet
traversed.

\begin{figure*}[h]
  {\relsize{-1}{
  \begin{center}
    \setlength{\tabcolsep}{1pt}
    \begin{tabular}{cc|c|cc}
      \multicolumn{2}{c}{\landscape}
      &~&
      \multicolumn{2}{c}{\island}
      \\
      \hline
      \includegraphics[width=.48\columnwidth]{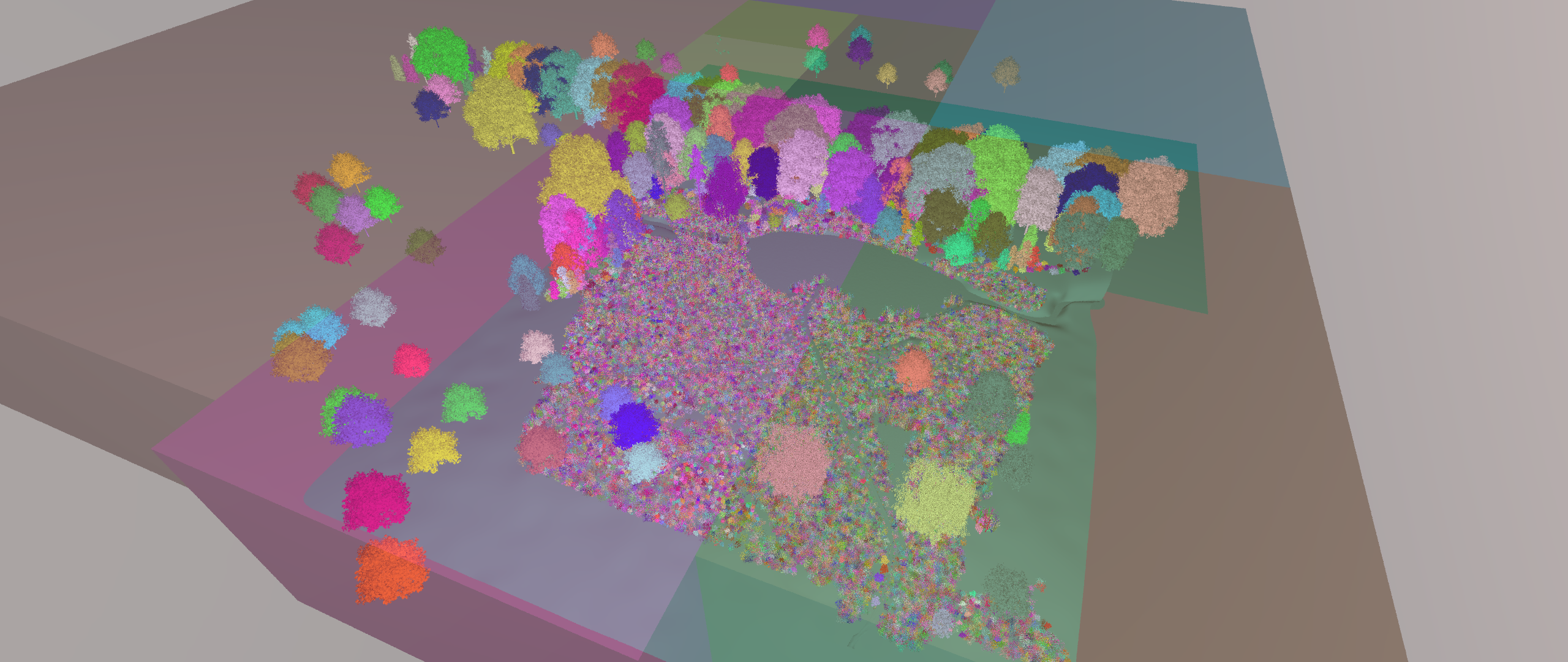}
      &
      \includegraphics[width=.48\columnwidth]{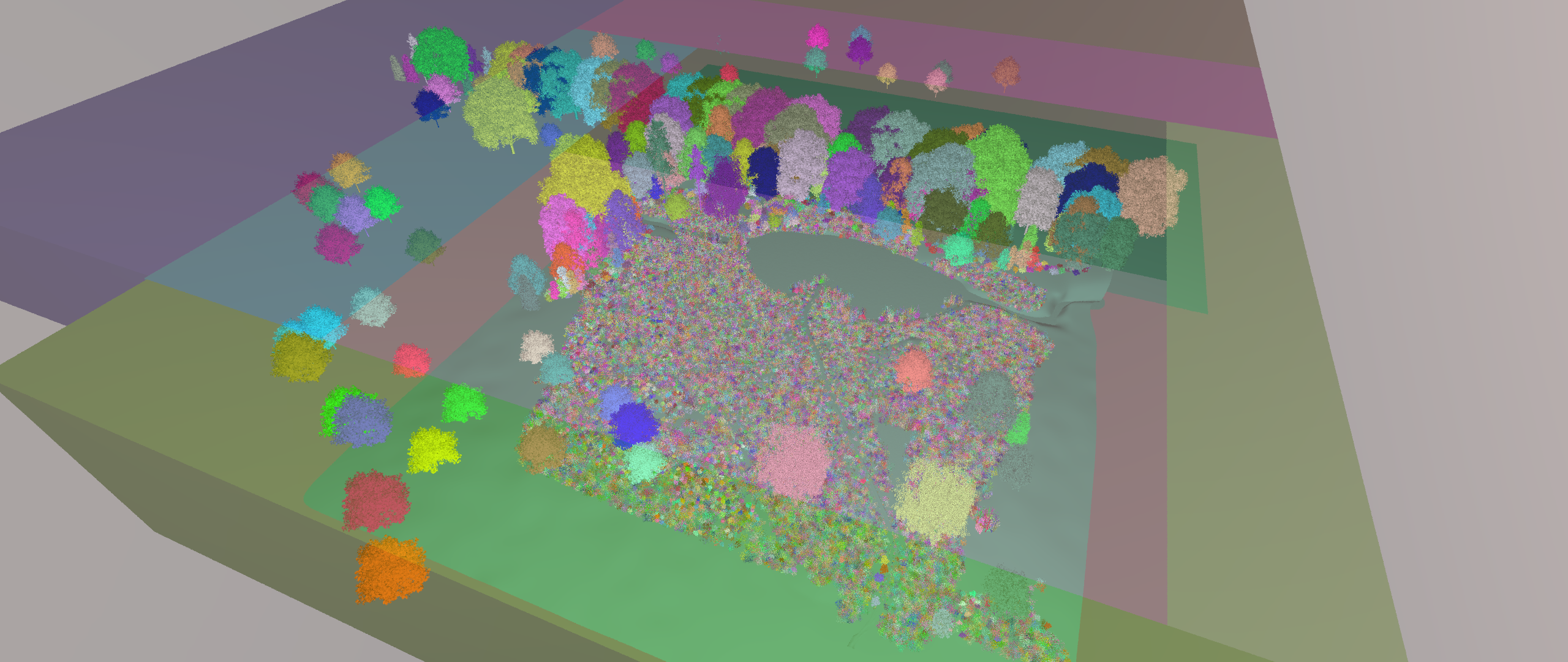}
      &
      ~
      &
      \includegraphics[width=.48\columnwidth]{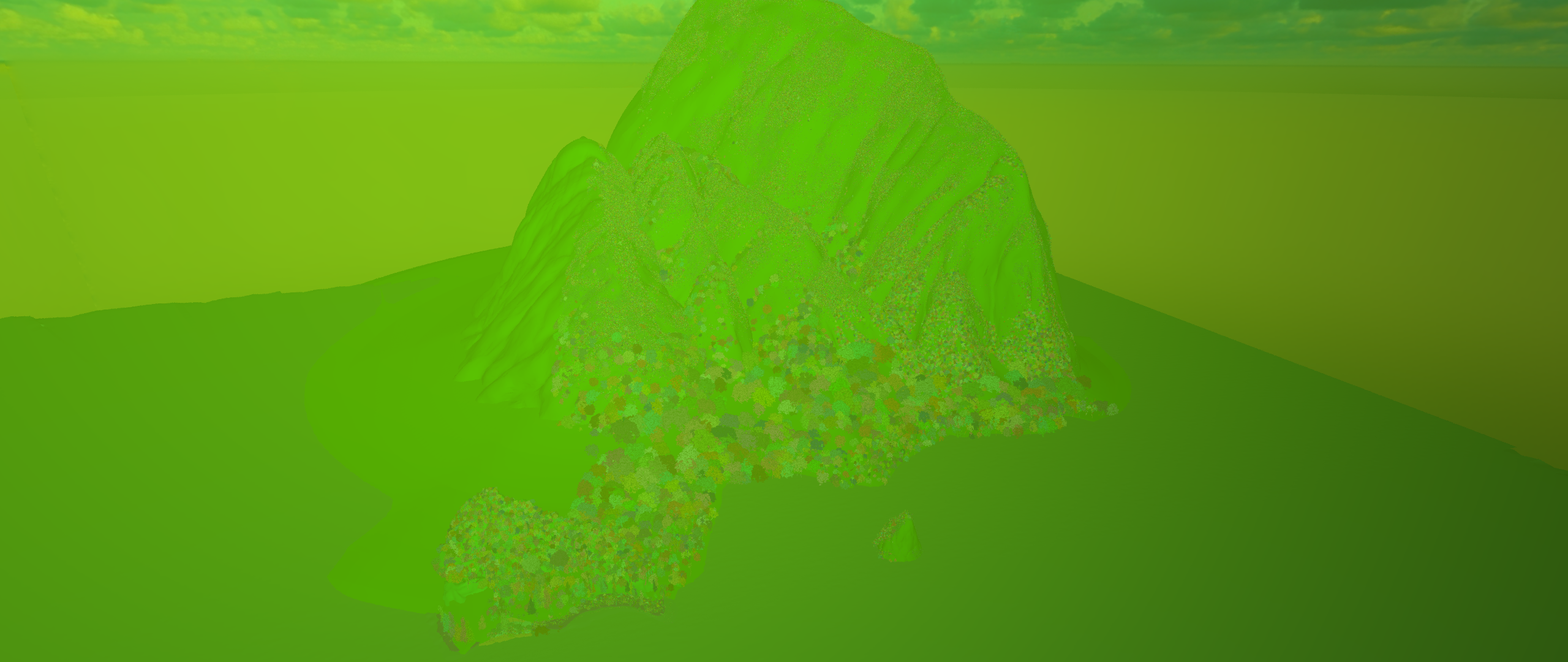}
      &
      \includegraphics[width=.48\columnwidth]{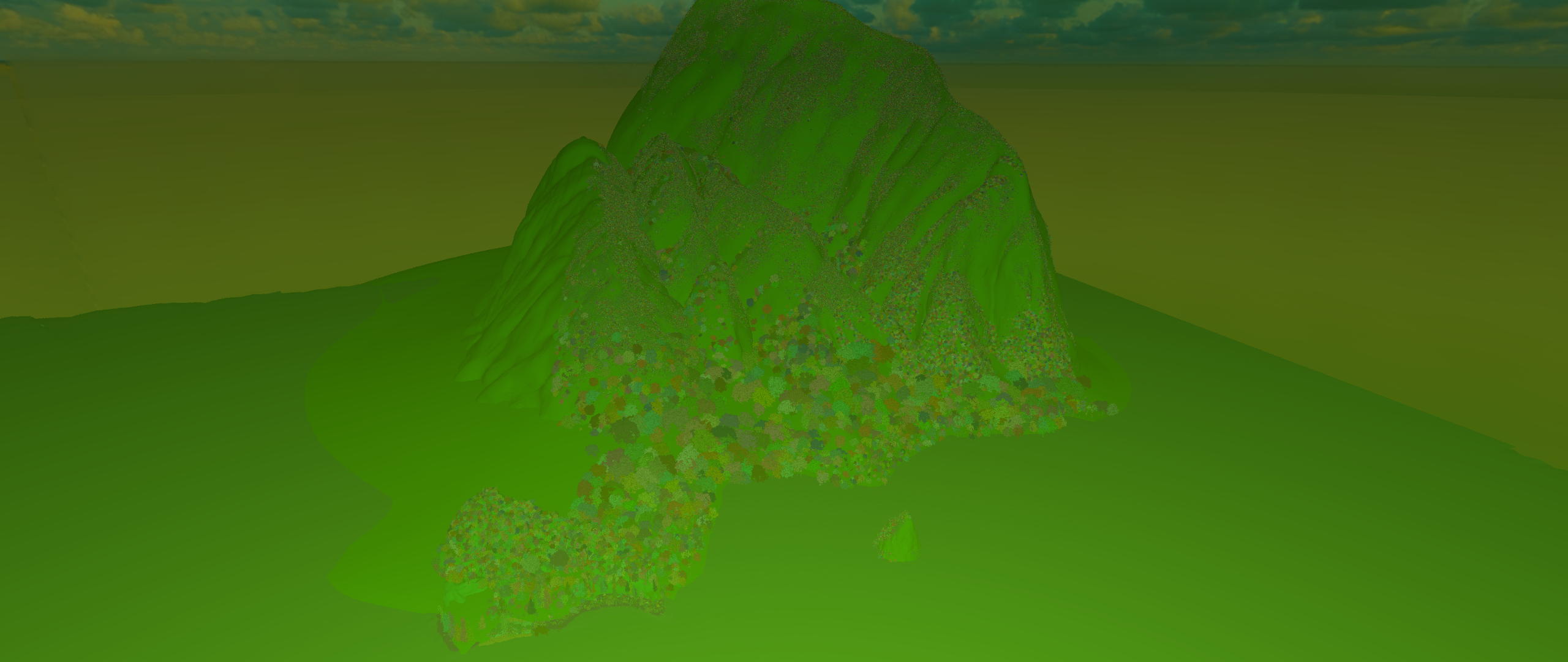}
      \\
      spatial (median) & spatial (w/ cost fct)
      & ~ &
      spatial (median) & spatial (w/ cost fct)
      \\
      \hline
      \includegraphics[width=.48\columnwidth]{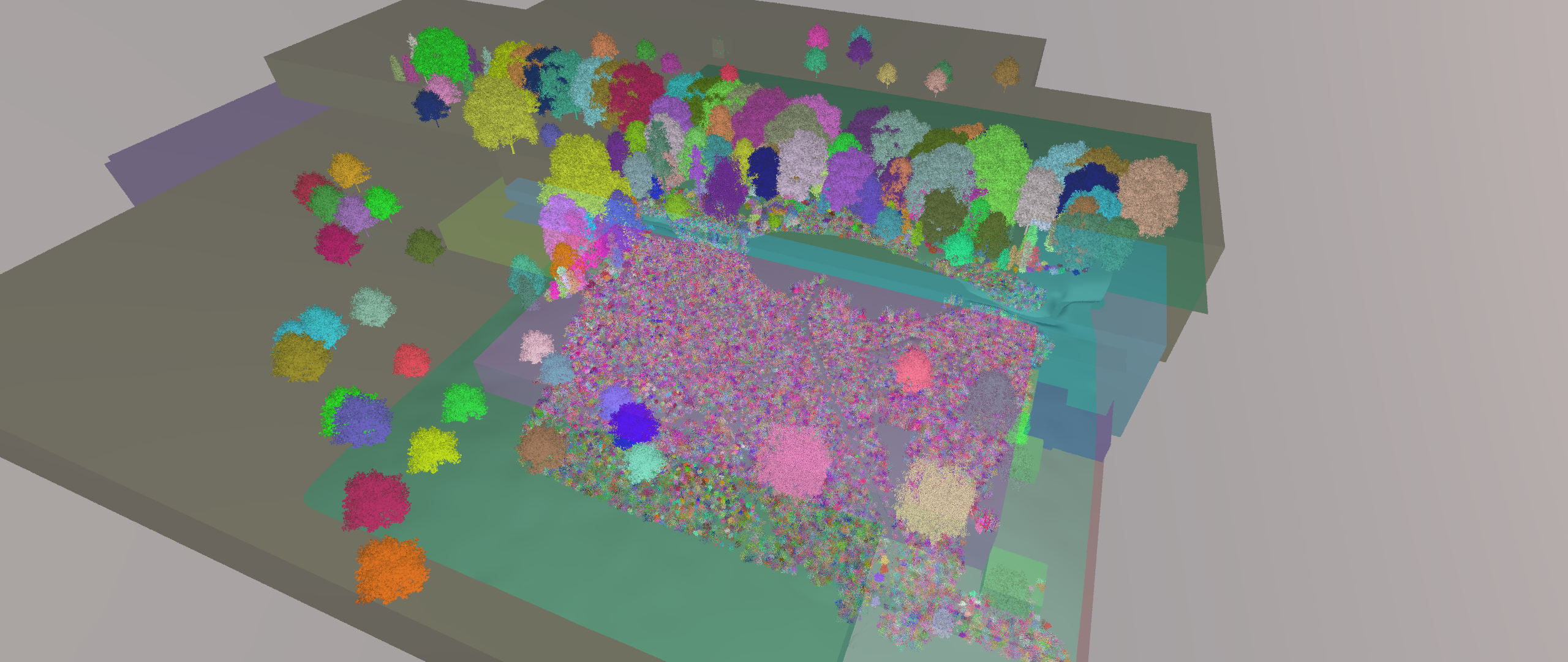}
      &
      \includegraphics[width=.48\columnwidth]{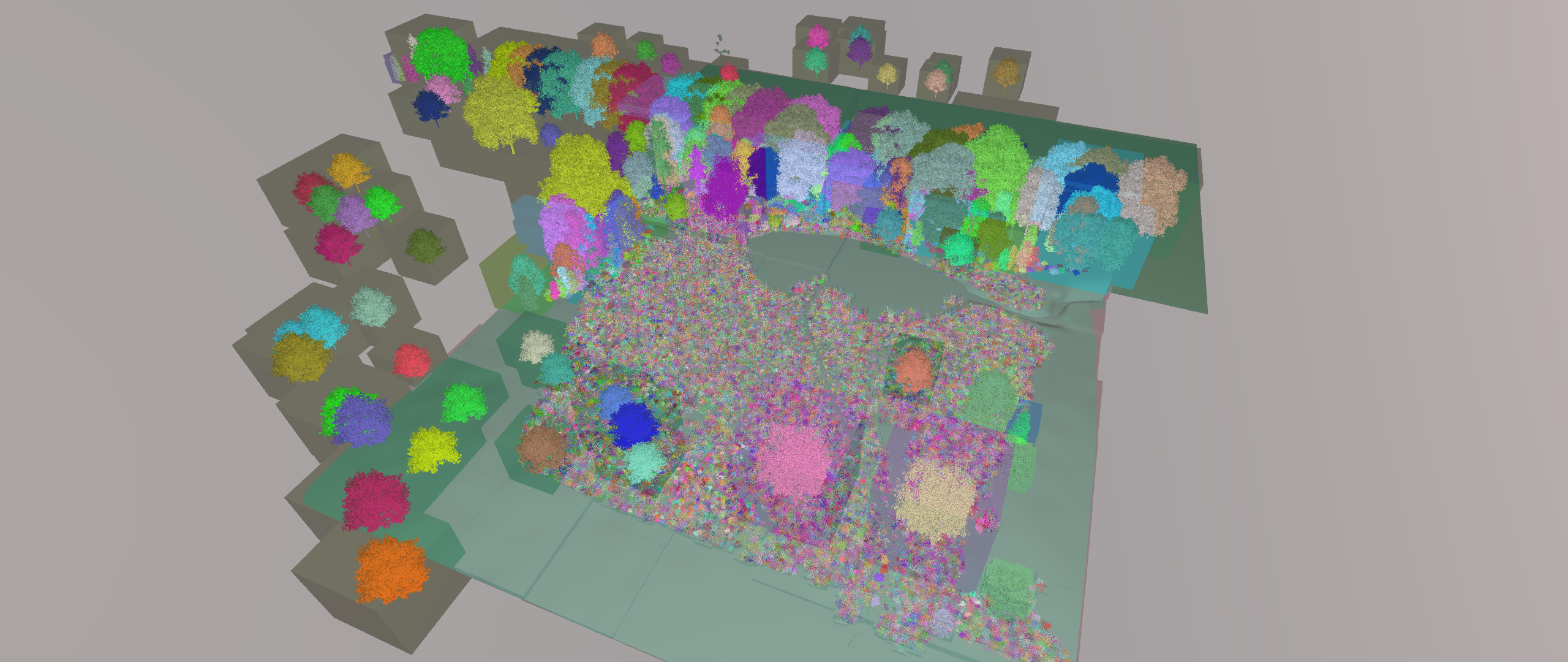}
      &
      ~
      &
      \includegraphics[width=.48\columnwidth]{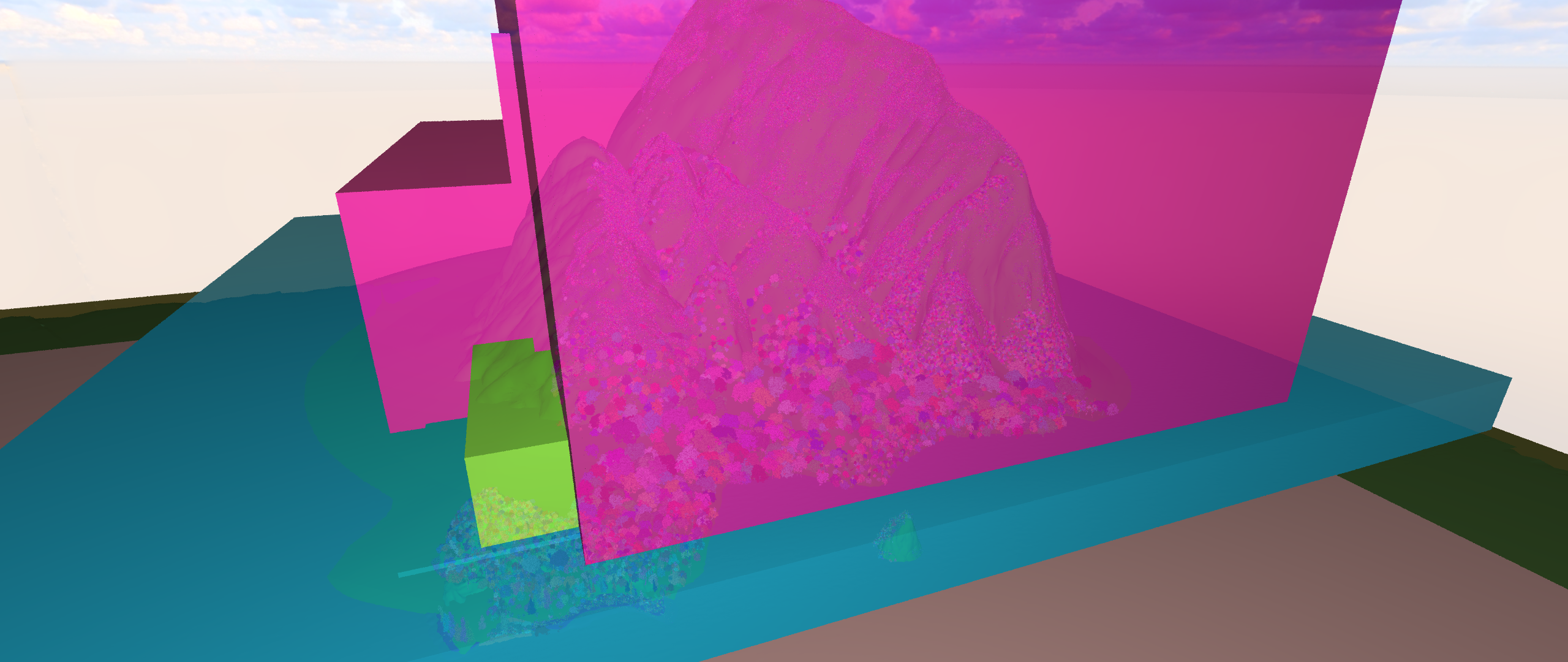}
      &
      \includegraphics[width=.48\columnwidth]{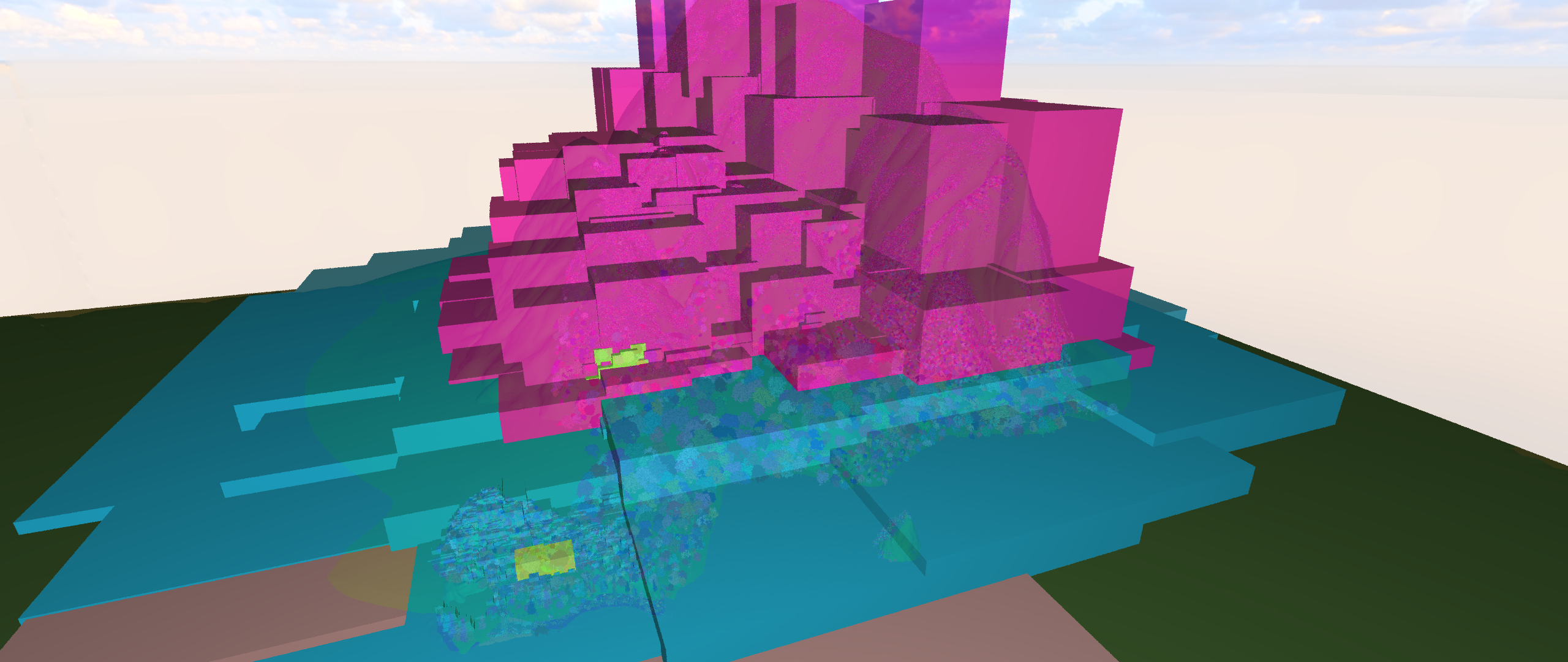}
      \\
      object (naive) & object (+ proxies)
      & ~ &
      object (naive) & object (+ proxies)
      \\
      \hline
      \includegraphics[width=.48\columnwidth]{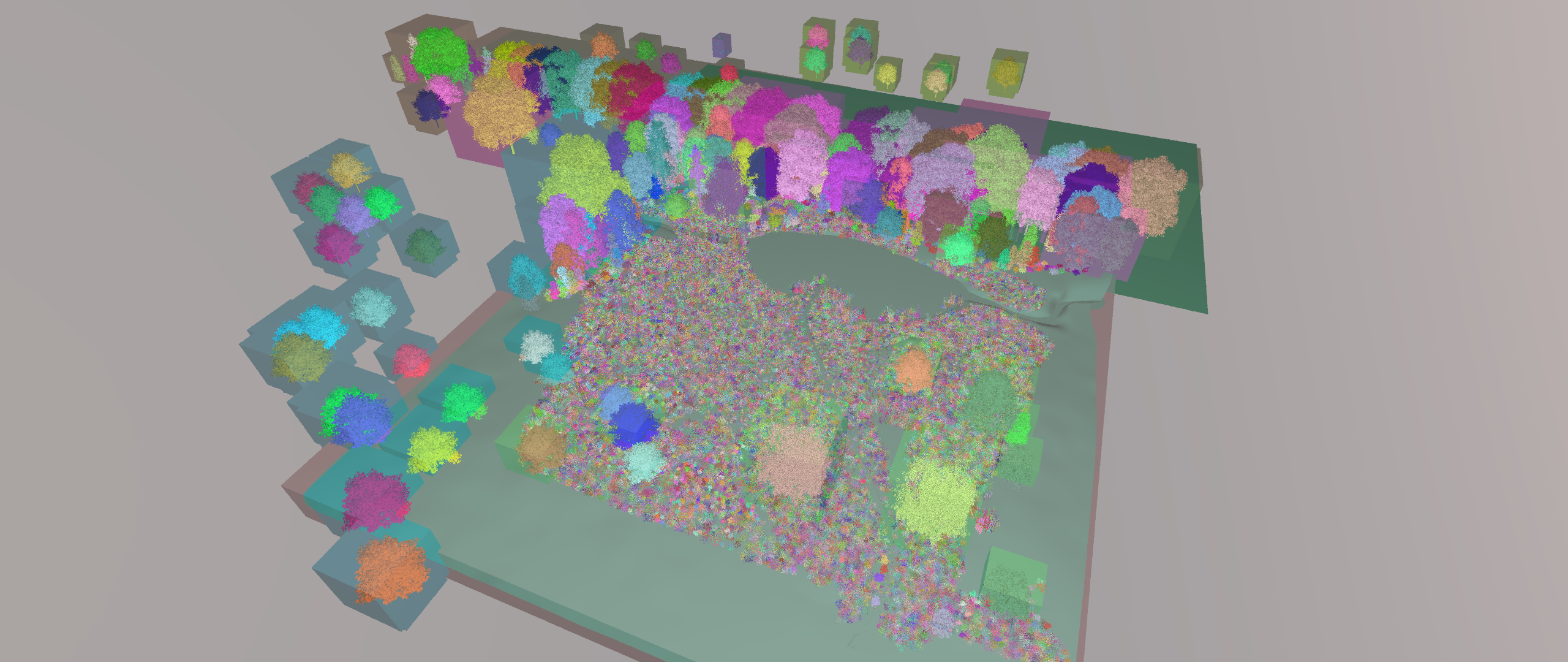}
      &
      \includegraphics[width=.48\columnwidth]{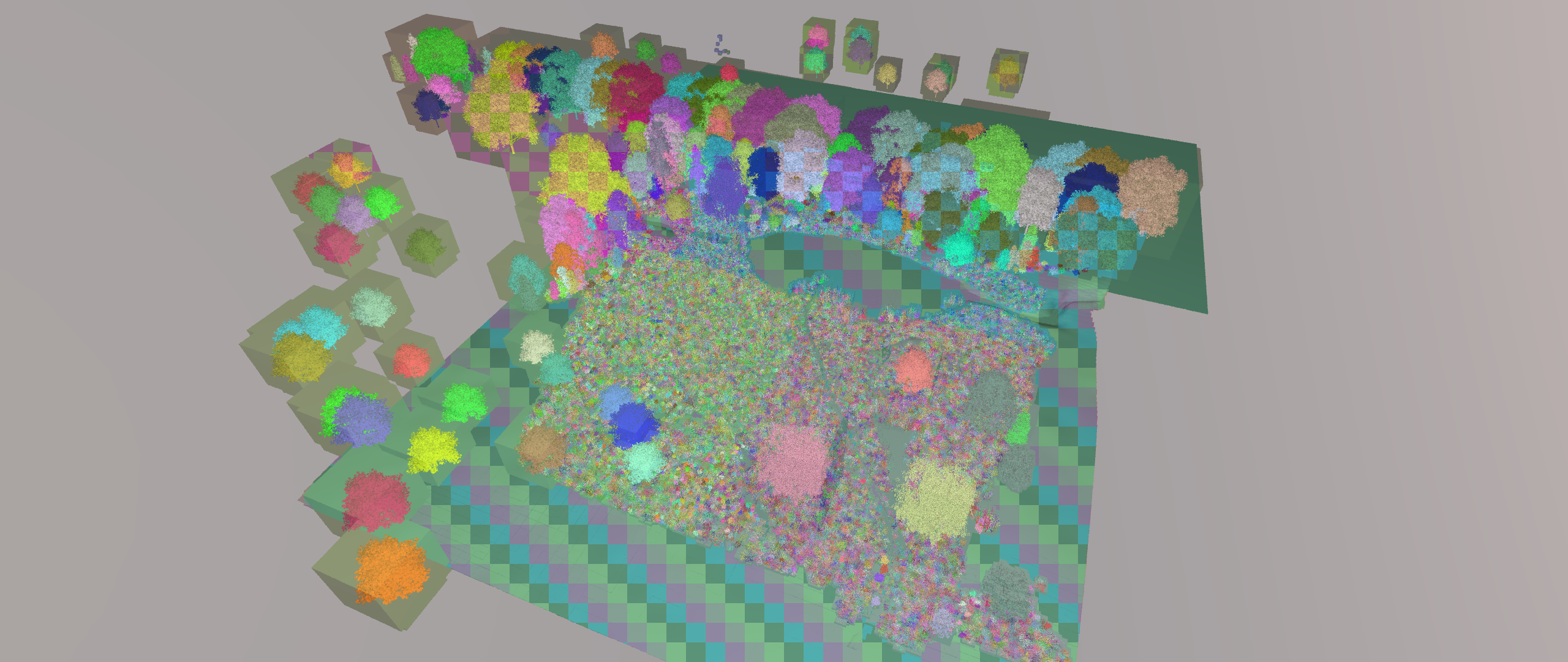}
      &
      ~
      &
      \includegraphics[width=.48\columnwidth]{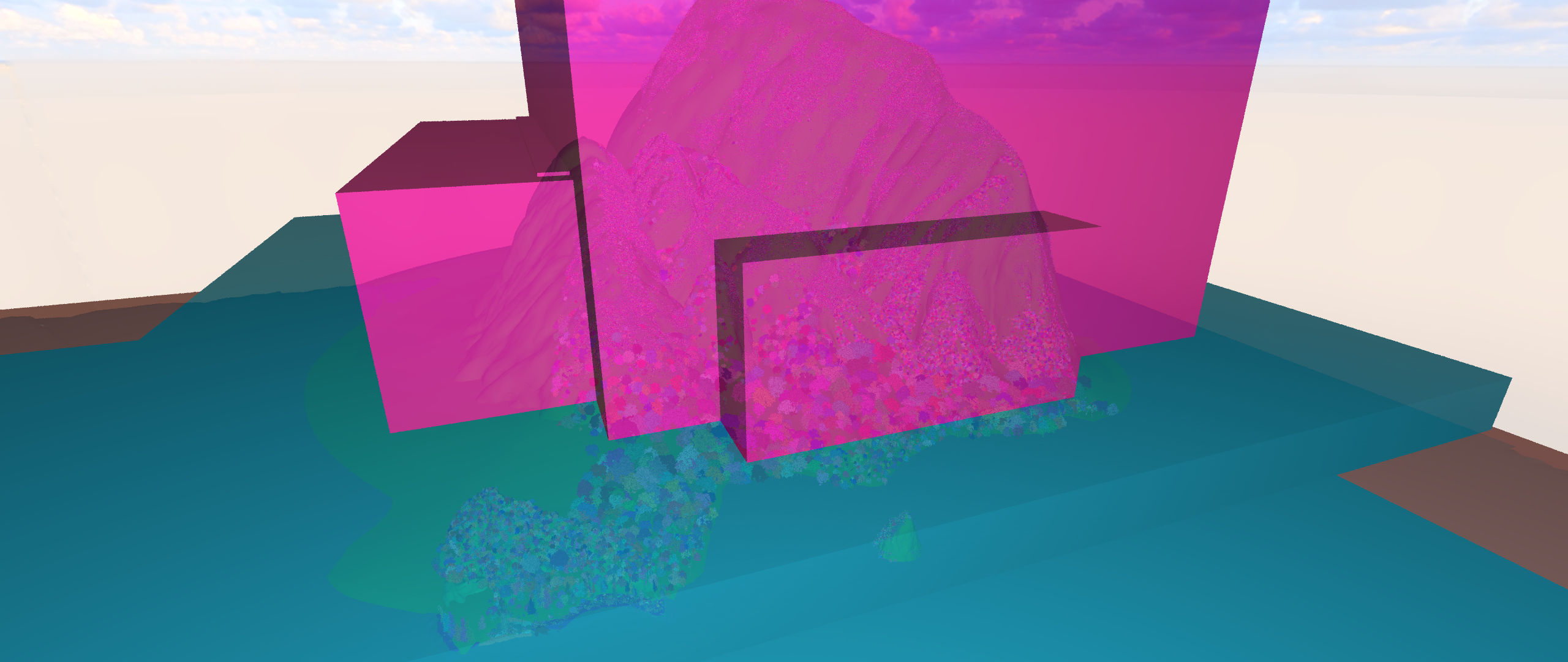}
      &
      \includegraphics[width=.48\columnwidth]{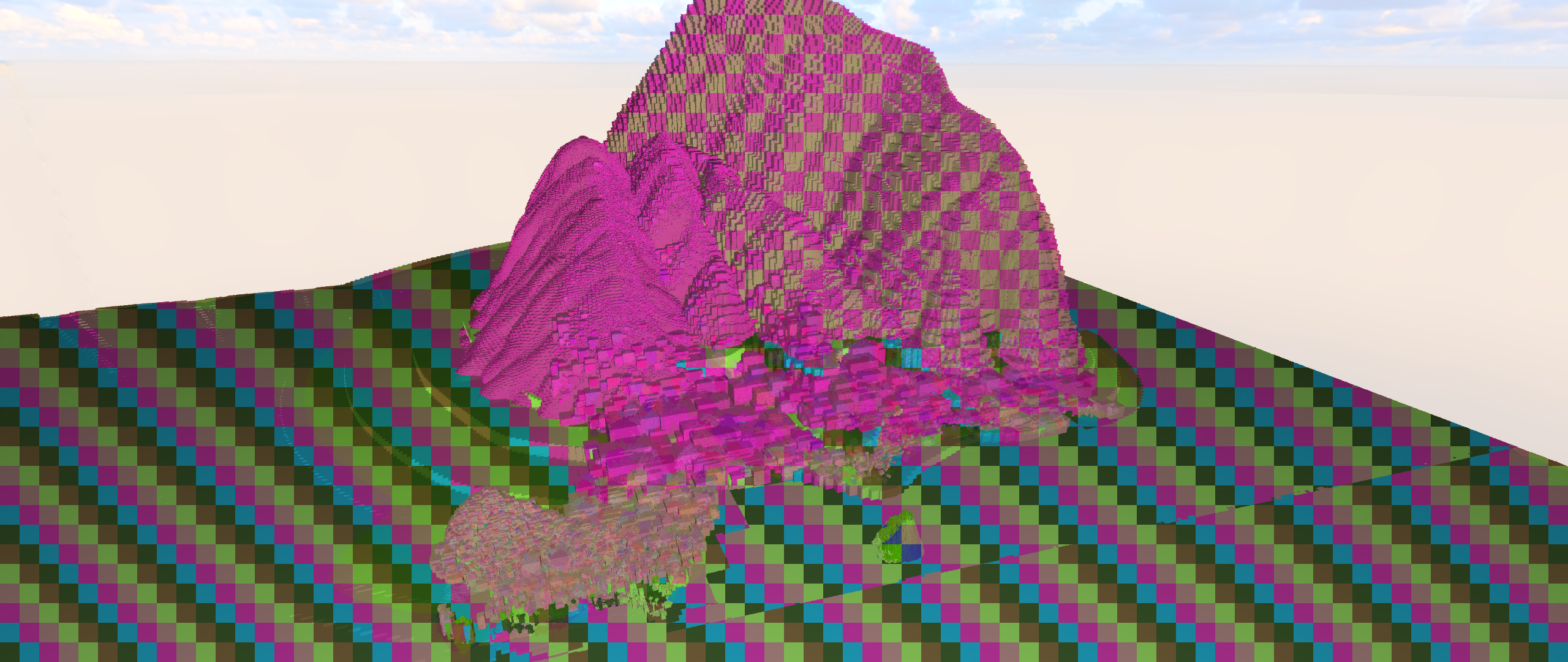}
      \\
      bvh-like & best (w/ some replication)
      & ~ &
      bvh-like & best (w/ some replication)
      \\
    \end{tabular}
  \end{center}}}
  \vspace{-2.99ex}
  \caption{\label{fig:proxies-per-model} Visual depiction of the model
    partitions that our sample partitioners produce for both
    \landscape and \island, with the boxes showing the used proxies'
    bounding boxes, and their color encoding the rank they are on (proxies with same
    color are on the same rank). Proxies
    with a checkerboard pattern mean that this content is owned on more than one rank.
    \vspace{-2.4ex} }
\end{figure*}

\subsubsection{Intersection local geometry.}
For a newly spawned secondary ray, even front-to-back traversal does
not guarantee that the spawning rank will get picked first; yet at
some time that ray would certainly get sent to that node. We always
first trace each ray on their generating ranks, then tag it as having
already visited this rank. In particular for shadow rays there is a
good chance that this new ray can actually find an occluder on that
same node, and never has to leave that rank at all.

\subsubsection{Generalizing proxies.}
Creating exactly one proxy per instance 
can lead to  very large proxies for some
objects, which in turn would require lots of rays to be sent to that
node. We observe that this is very similar to problems that have recently
been investigated for BVH traversals, and in particular point to
techniques like \emph{spatial
  splits}
and \emph{braiding}.
Both of these work by  representing spatially large objects
in a BVH 
through more than
one box, with the entirety of these boxes covering the object more
tightly than one single box. We use this to conservatively but tightly
represent spatially large instances with multiple smaller boxes (see
Figure~\ref{fig:sketch}d), which allows rays to pass around
some geometry whose proxies they would otherwise have hit. Rays can now encounter
multiple proxies of the same object, but the previous
paragraphs' techniques skip these, so this is
OK.
Ultimately this means that proxies no longer represent any
particular instance, but just \emph{a region of space that a given
  node has content for}. I.e., we can have one proxy represent more
than one instance, or use multiple proxies for the same instance, etc.

\subsubsection{Replicating certain geometry.}
Just like proxies are not tied to any particular instance, so we can
generalize the concept of who owns the content behind one
proxy. Though we did initially assign exactly one rank to every
instance, for some spatially large (and thus, likely to get traversed)
yet not memory intensive object(s) we might also want to replicate
this object to more than one node, such that rays already on that node
would have it available without needing to travel to the node owning
it. With our proxies, we can easily do that, by allowing proxies to
specify that whatever it may represent, it can be found on more than
one node.  The traversal logic above does not
change at all:
when we trace a ray to find the next proxy we simply reject all
proxies for which the ray has been to \emph{any} of the nodes listed
in that proxy.

\subsubsection{Proxy-guided primary ray generation.}
\label{sec:primary-rays}
Above we have
 described that secondary rays should always be traced on the
node that spawned them, but for primary rays we can actually
\emph{choose} where to spawn them. In particular, we can use
our proxies to generate each path on exactly the node
that owns the closest proxy for the given pixel, thus maximizing the 
chance that this ray will find its first intersection on exactly the node 
it was generated on.

\subsubsection{Tracking which nodes a ray has already been on.}
\label{sec:replay-technique}
In all of the previous techniques we have made the implicit assumption
that a ray always known which node(s) it has already been on. 
For not too large a number of ranks this can 
be realized through a bit mask (with one bit per rank) that gets
attached to each ray. For a large number of ranks this would lead
to an explosion in ray size; however, this can be avoided by
what we call the ``replay'' technique: if each ray only ever
stores which rank it was generated on, then any node can later
re-compute the actual set of visited nodes by simply re-running the
above logic until it reaches itself.

\medskip
In combinations these technique provide a very effective \emph{operator} 
that---using only the proxies, the ray, and the ray's stored history---allows any node to robustly and efficiently
determine which other rank to forward that given ray to next; if this comes up empty that ray's distributed traversal is complete. 

\section{Partitioning}
\label{sec:partitioning}

This paper is not about one particular partitioning strategy; in fact,
we believe our method's greatest strength is its ability to express
and handle distributed content in a more general way. To demonstrate
this flexibility we implemented multiple different partitioners,
including both spatial, object space, and hybrid methods.  All our
variants work similarly in that we start by creating one \emph{part}
containing the whole scene, then iteratively take the
respectively largest part, and split that into two. For objects with
more than one instance we use the individual instances of that object,
for those with only one instance we follow
Zellmann~\shortcite{zellmann} and break that object into its
constituent meshes.

\emph{Spatial} partitioning starts with an initial \emph{domain} set
to the scene's bounding box, and in each split creates two
non-overlapping halves, then checks which objects overlap each half's
domain. Again following Zellmann, after each step each side's domain
gets shrunk to the content it contains, if possible. For deciding \emph{where} 
to split we implemented two methods: \code{spatial-simple} 
splits each domain at its spatial median; \code{spatial-sah} uses a
cost function to pick one among $3\times 7$ equidistant candidate
splits. As cost function we first compute how many unique meshes,
triangles, vertices, texels, etc., each rank has, then weigh these with
an estimated memory cost for each such item; the final cost of a split
then is a 50:50 blend between traditional SAH and the sum of these
memory estimates. As proxies we eventually use exactly those domain
boxes, at which point out method can handle this data. This is, in
fact, a powerful finding: the techniques we propose are not the exact
\emph{opposite} of spatial subdivision, but a \emph{generalization}:
though we can do more, spatial subdivision can be represented just as
well, and some of our optimizations can even be back-ported into
 spatial subdivisions.

\emph{Object} partitioning works on objects, not instances: all
instances of an object always go to the same rank. For each object we
create one box around all its instances, then use this box to sort
this object left or right of any candidate plane. Again we use
$3\times 7$ planes, and our cost function to pick the best one. For
computing the proxies we implemented two methods: \code{object-naive}
uses the same boxes as used for partitioning; \code{object-proxies}
creates one proxy for every instance, and by default 64 smaller
proxies for non-instanced meshes. The latter we compute just like with
braiding, by performing a number of BVH build steps on each such mesh.

\emph{Hybrid} partitioning combines both techniques: we partition
based on instances, not objects---so \emph{some} instances can get
replicated, if the cost function so chooses. Otherwise partitioning is
similar to object space, using the same cost function. In
\code{bvh-style} we use one box per instance respectively non-instanced mesh; these are
also used as proxies. In our currently \code{best} partitioner we
first---\emph{before} partitioning---split large objects into multiple
boxes, then partition these. This means that some objects can now get
assigned to more than one rank if the cost function so chooses.
The partitions and proxies resulting from these
strategies are shown in Figure~\ref{fig:proxies-per-model}.

\begin{figure*}[h]
  {\relsize{-1}{
  \begin{center}
    \setlength{\tabcolsep}{1pt}
    \begin{tabular}{ccc}
      \includegraphics[width=.325\textwidth]{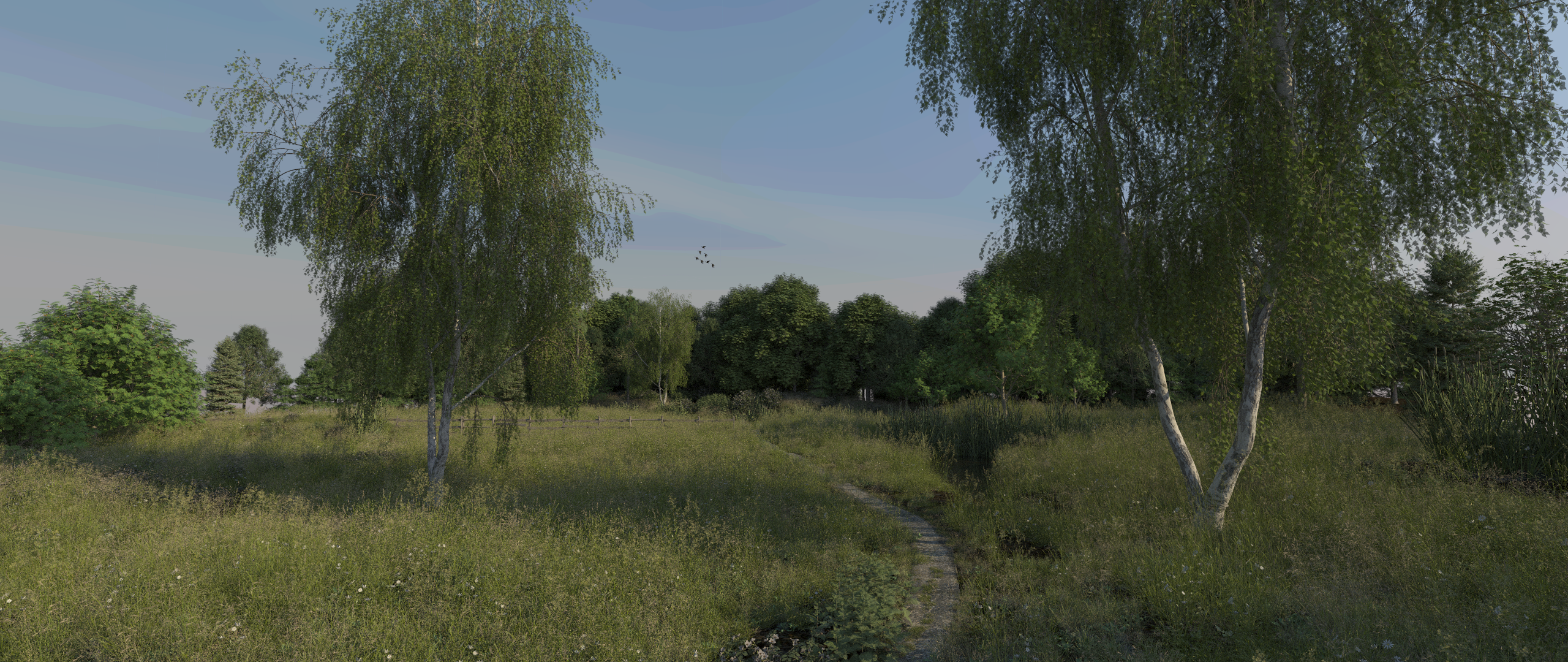}
      &
      \includegraphics[width=.325\textwidth]{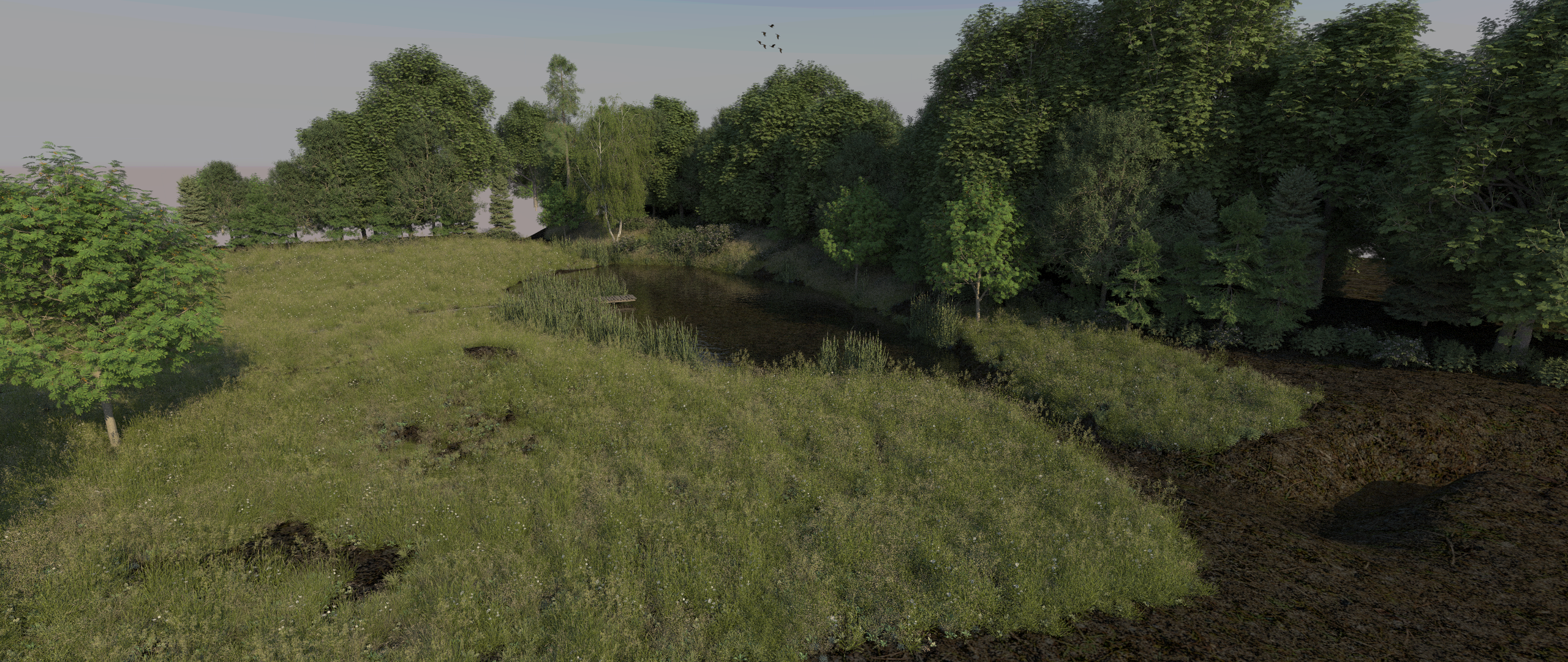}
      &
      \includegraphics[width=.325\textwidth]{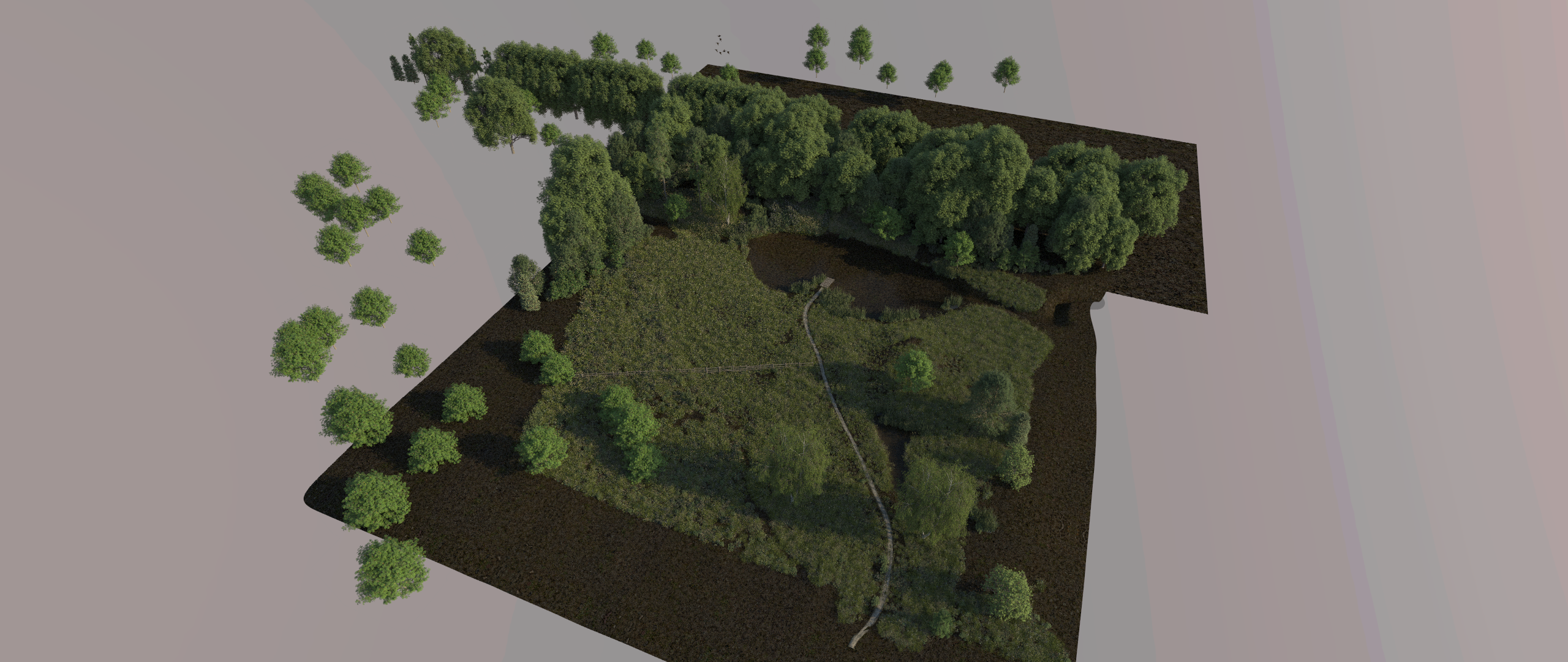}
      \\
      \multicolumn{3}{c}{\landscape model: \quad max GPU memory usage (on 8 ranks, including non-model data like ray queues, frame buffers, etc.) 3.7~GB (ours) vs 4.8~GB (spatial)}
      \\
      \emph{path}: 6.1~fps (ours) \quad 5.2~fps (spatial) &
      \emph{view-3}: 9.1~fps (ours) \quad 4.7~fps (spatial) &
      \emph{top}: 13.3~fps (ours) \quad 10.1~fps (spatial)
      \\
      \includegraphics[width=.325\textwidth]{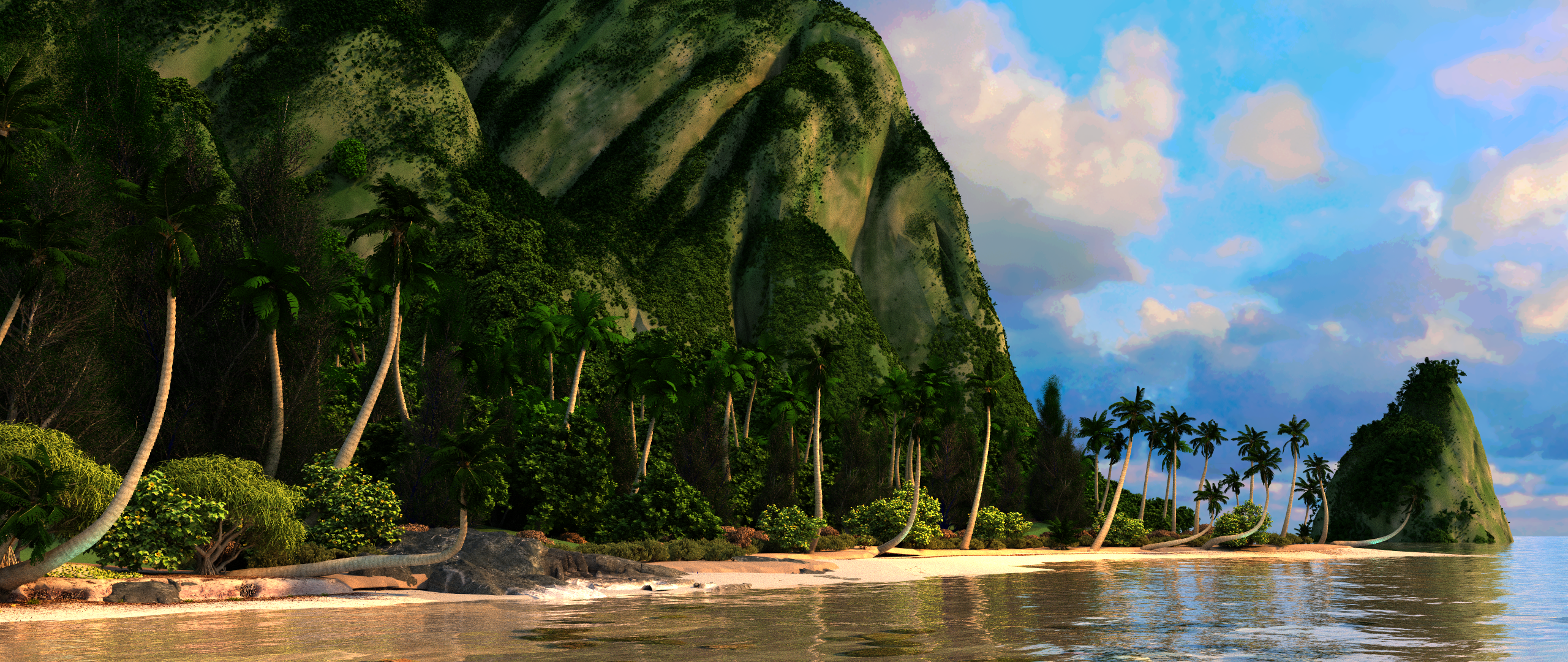}
      &
      \includegraphics[width=.325\textwidth]{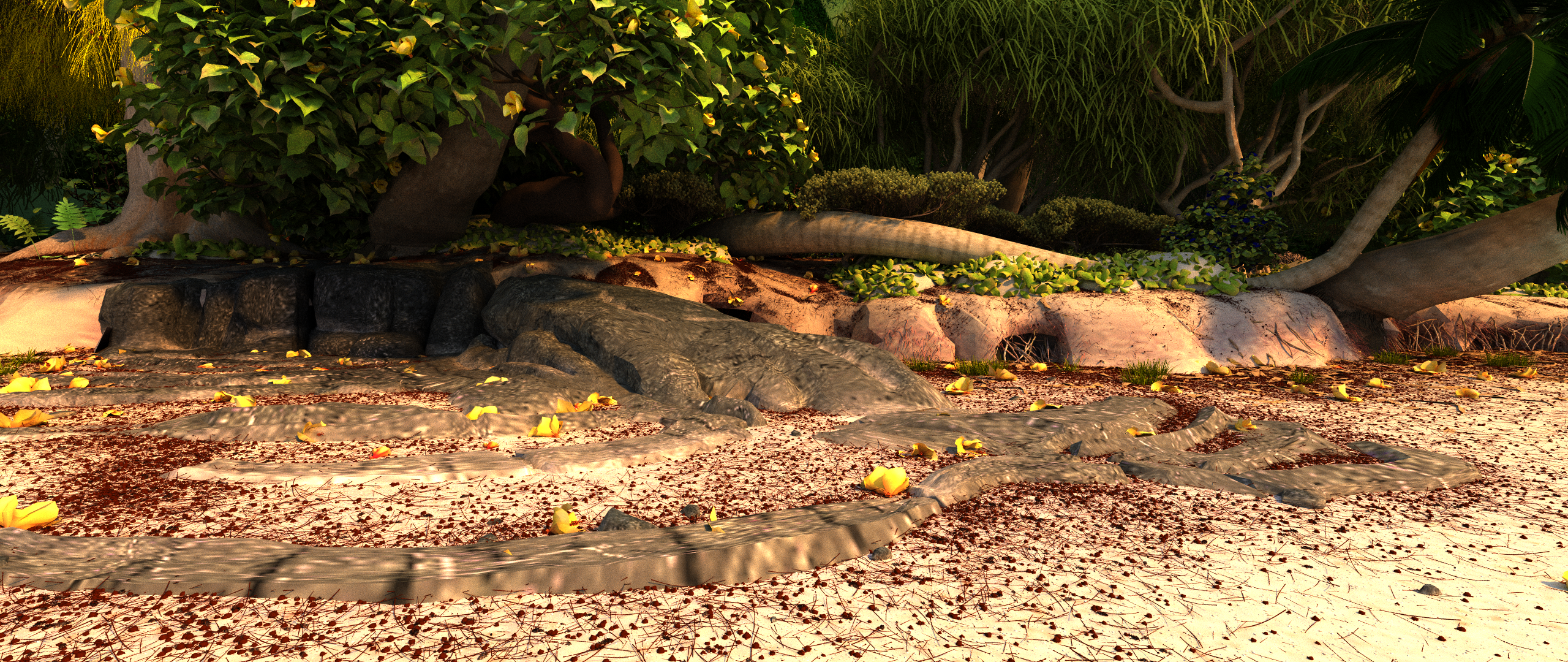}
      &
      \includegraphics[width=.325\textwidth]{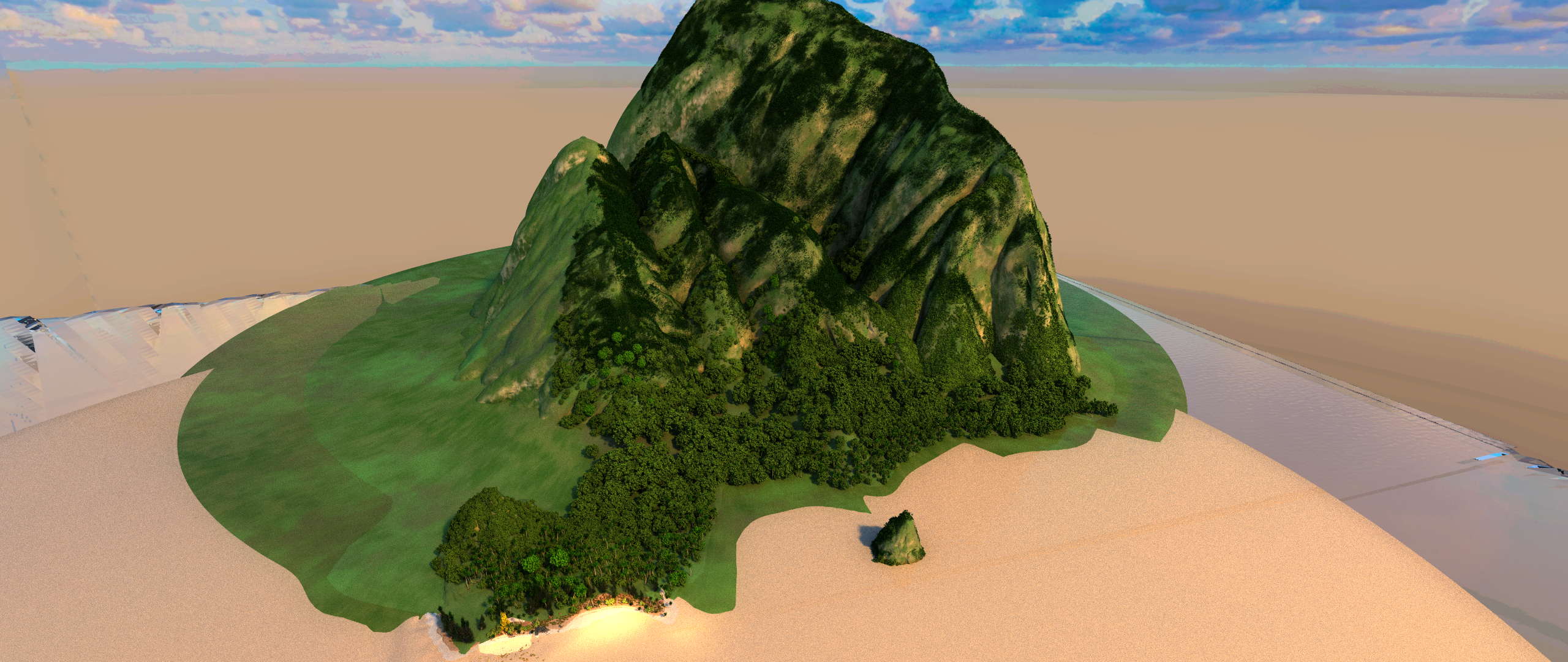}
      \\
      \multicolumn{3}{c}{\island model: \quad max GPU memory usage (on 8 ranks, including non-model data like ray queues, frame buffers, etc.): 25~GB (ours) vs 48~GB (spatial)}
      \\
      \emph{default}: 7.9~fps (ours) \quad (3.0\textsuperscript{$\star$})~fps (spatial) & 
      \emph{beach}: 3.6~fps (ours) \quad (2.8\textsuperscript{$\star$})~fps (spatial) &
      \emph{overview}: 11.1~fps (ours) \quad (3.7\textsuperscript{$\star$})~fps (spatial)
      \\
    \end{tabular}
  \end{center}}}
  \vspace{-2.8ex}
  \caption{\label{fig:eval-views}\label{fig:views-per-model}
    \label{fig:results}
    Views used for our
    evaluation, and render performance for these views.
    ($2560\times 1080$, one path/pixel, 4 workers@$2\times$RTX~8000, 10-GigE Ethernet).
    \\(\textsuperscript{$\star$}: \emph{spatial} can render this model only if we upgrade the first rank to 128~GBs of RAM, and even then requires significant swapping during scene setup)
    \vspace{-3.2ex}
}
\end{figure*}

\section{Implementation}
\label{sec:full-system}

The core contribution of this paper is not any one implementation, but
the general concepts of looking beyond purely spatial partitioning,
the proxies, and the specific techniques for the data-distributed
traversal operating on these proxies. Nevertheless, to prove that
these concepts do in fact work we also developed a 
data-distributed GPU path tracer that uses these concepts.

For communication we use a CUDA-aware
build of OpenMPI 4.1.2, which means that we can directly pass device
addresses to MPI, which then copies data as required. With
better hardware this would also allow RDMA communication between
GPUs and network devices, but on our low-end setup this is not the case.
The renderer uses a wavefront-design, with all shading, compaction,
etc., done in CUDA 11.4, and all tracing done using OptiX 7.4.

That
same renderer can also be recompiled to a CPU-only version that uses Embree
and TBB. The same concepts work there as well, but for our model the texturing in particular
is an issue on that setup. A detailed comparison is beyond the scope of this paper.

\subsection{Rays, Paths, Hits, and Ownership Masks}
\label{sec:re-trace}
  
Our framework builds on  small, self-contained
\emph{path nodes} that can be forwarded across the network.  Each path
contains a ray origin and direction, a throughput
value~\cite{rt-gems-reference-path-tracer}, and the pixel ID to which
it belongs, plus some bits to indicate whether a ray is  a shadow ray, in
medium, etc.  To track already visited nodes we use a bit-field of
either 8 or 64 bits depending on number of ranks; for more ranks we
would use the technique described in
Section~\ref{sec:replay-technique}. We use half
precision for ray direction and throughput, and encode all bits and
pixel ID in a 32-bit integer.

For the currently closest intersection we only store the distance in
\code{ray.tmax}, plus the node mask of the geometry that was hit. This
means paths have to be re-traced for shading, but this is cheaper than
sending hit information across the network. We could also have stored
the node ID that produced the hit; but using a node mask is better: if
the ray were to need forwarding and later terminates on another node,
having the node mask of the hit allows this other rank to check if it,
too, happens to have that data, thus allowing the ray to be shaded
there without it having to be sent back.

Using this encoding, the path struct is a mere 36 bytes in size. We
refer to that same structure as either \emph{ray} or a \emph{path}
depending on context, but always mean that same struct. Paths always
get generated and shaded in wavefronts; between two such shading
operations each wavefront goes through a distributed traversal until
every path is on the node it can be shaded on.

\subsection{Distributed Path Tracing}

For the forwarding logic we use an OptiX acceleration
structure built over the proxies, with an intersection program that
rejects all proxies whose ownership mask lists any node that the ray
has already been on.
If any next proxy was found we use its bit-mask to pseudo-randomly pick one
of the ranks listed in that mask. Otherwise, the ray is done
traversing, and can go to shading. In that case we check if that ray
can be shaded on the current node, and if not, pseudo-randomly pick one
of the ranks listed in its hit mask.

\subsubsection{Wavefront Ray Traversal.} Our core operation
is to take one wavefront of rays, and trace these---across
nodes---until each ray has terminated traversal, and is on a node
where it can be shaded.  We call this the \emph{distributed ray
  traversal}, and it proceeds in three stages:
We first launch an OptiX program that traces each ray into
the rank's local geometry. This uses an \code{anyhit} program to do
alpha texturing, and a \code{closest hit} program that updates the
ray's \code{tmax} and hit mask if a closer intersection is found. The
program then updates the ray's \code{alreadyVisited} mask, traces it
into the proxy acceleration structure, and determines which rank that
ray needs to be sent to next as described above.
We then run a CUDA compaction kernel that rearranges all rays such
that those that can be shaded locally go to one place, and those that
need forwarding go to another, with the latter sorted by the rank they need
to go to.

Once the rays are thus arranged all ranks collaboratively
execute an MPI \code{Allgetherv} to exchange how many rays each rank
wants to send to any other node, followed by an MPI \code{All2all} that
moves the rays to their respective destinations.  These three stages
get repeated until no more rays need exchanging, at which point every
rank has a  wavefront of rays ready to be shaded on that rank.

\subsubsection{Wavefront Shading and Secondary Ray Generation}

After a wavefront has been traced to completion each rank locally
shades its rays.  Shadow rays that terminated traversal on the current
rank check if that shadow ray did find an occluder, and if so,  get
discarded; those that didn't atomically add their throughput value
into the rank's frame buffer. For non-shadow rays, those that
did not find an intersection get shaded by either background or
environment light, and get accumulated into the frame buffer.

For a non-shadow ray that did have a hit we first re-trace that ray
into the local geometry to re-compute the full hit and BRDF data
(Section~\ref{sec:re-trace}). We sample the BRDF to produce either a
reflected or refracted ray, modify the ray's throughput value
according to the sampled BRDF, and use rejection sampling to avoid
tracing rays with too low a throughput value. The secondary ray---if
not rejected---gets appended to the next step's wavefront queue.

Shading can also generate a shadow ray. To prevent possibly unlimited
growth of the ray queues we use repeated reservoir
sampling~\cite{reservoir-sampling} and importance sampling to always
choose at most one sample from possibly multiple different lights and
light types. Thus any pixel can have at most two rays active at any
time: one for the path itself, and one for its corresponding shadow
ray. Shadow rays first compute the pixel contribution they would have
if not occluded, then store that value in their throughput field, and
set a bit in the path that flags this as a shadow ray.

\subsubsection{Proxy-Guided Primary Ray Generation.}

For primary rays we use the technique described in
Section~\ref{sec:primary-rays}: each rank generates
\emph{every} primary ray and traces it into the proxy acceleration
structure (which on modern hardware is very cheap). This ray then
picks a \emph{primary owner} based on the bits
of the closest proxy, and all but one rank will then discard this ray.
For rays that hit proxies stored on more than one rank we use the
pixel ID as a tie-breaker, which in the pseudo-color images in
Figure~\ref{fig:results} can be observed as an checkerboard pattern on
those objects that the partitioner chose to replicate. 
This is intentional, and allows the
work to get shared by all the nodes that have the data.

\subsection{Merging Ranks' Partial Frame Buffers}
\label{sec:partial-fbs}

Irrespective of which rank a path was generated on, it---and the
secondary rays it may spawn---can terminate on any other node; so,
every pixel can get contributed to by any rank. One way to
handle that is to send every shading contributions back to the
rank that generated the path; but that is expensive.
Instead, we have each rank maintain a full frame buffer for all the
image contributions computed on that rank.  These
partial-sum frame buffers eventually need to get added for the
final image. For this we  use what in visualization is known as
\emph{parallel direct send} compositing~\cite{parallel-direct-send}:
each rank is responsible for one part of the final
frame buffer, and gets
other ranks' contributions from those ranks.
Each rank then adds up the parts it received,
performs tone mapping, and sends the final RGBA pixels
to the master.

\section{Evaluation}

Using the implementation described in the previous
sections we can now evaluate how well these techniques work.
Since this is the only large production content that is
publicly available we focus primarily on the
\emph{island} model, but also include the PBRT
\emph{landscape} for reference. For \island we enable the
\code{isMountainA/xgLowGrowth}, and tessellate all \code{curves} geometry
into 3 linear segments per curve segment.
For the original model's PTex
textures~\cite{ptex} we  perform a baking step that creates a
per-mesh texture with $8\times 8$ texels per quad, with properly created
texture coordinates added to each vertex.

For hardware we use what is called ``Beowulf'' cluster of, in our
case, five similar networked PCs, using one as head
node, and four as workers. Each worker is equipped with 64~GBs of
RAM, and with two 48~GB RTX~8000 cards; the master only runs the
display. For networking we use a commodity at-home 10Gig-E
Ethernet, which is well below what modern data center hardware
can provide (e.g., a Mellanox ConnectX-7 is rated at 400~GBit/s, vs
our 10~GBit/s). Better interconnects also allow RDMA transfers,
which our setup does not. Using such low-end network may look
counter-intuitive, but is useful to establish a baseline,
and forces us to always focus on the main problem: bandwidth.

\subsection{Maximum GPU Memory Usage}
\label{sec:max-part-size}

The ultimate rationale behind data-parallel rendering is to reduce the
amount of data per node until it fits per-node memory. 
As such, we first evaluated how well different
partitioning strategies performed in reducing per-rank memory.  To do
this we took the \island model, split it into a varying number of $N$
parts (from $N=1..128$), and logged the size of the respectively
largest part---which is the part that would most exceed the
memory budget. To measure this we used the cost function
described above; this doesn't include rendering related data like ray
queues, etc., and is an approximation even for model data; but is
hardware agnostic and  easy to compute.

\begin{figure}[h]
  \vspace*{-3.5ex}
  \adjustbox{trim={.001\width} {.01\height} {0.001\width} {.6\height},clip}{\includegraphics[width=.999\columnwidth]{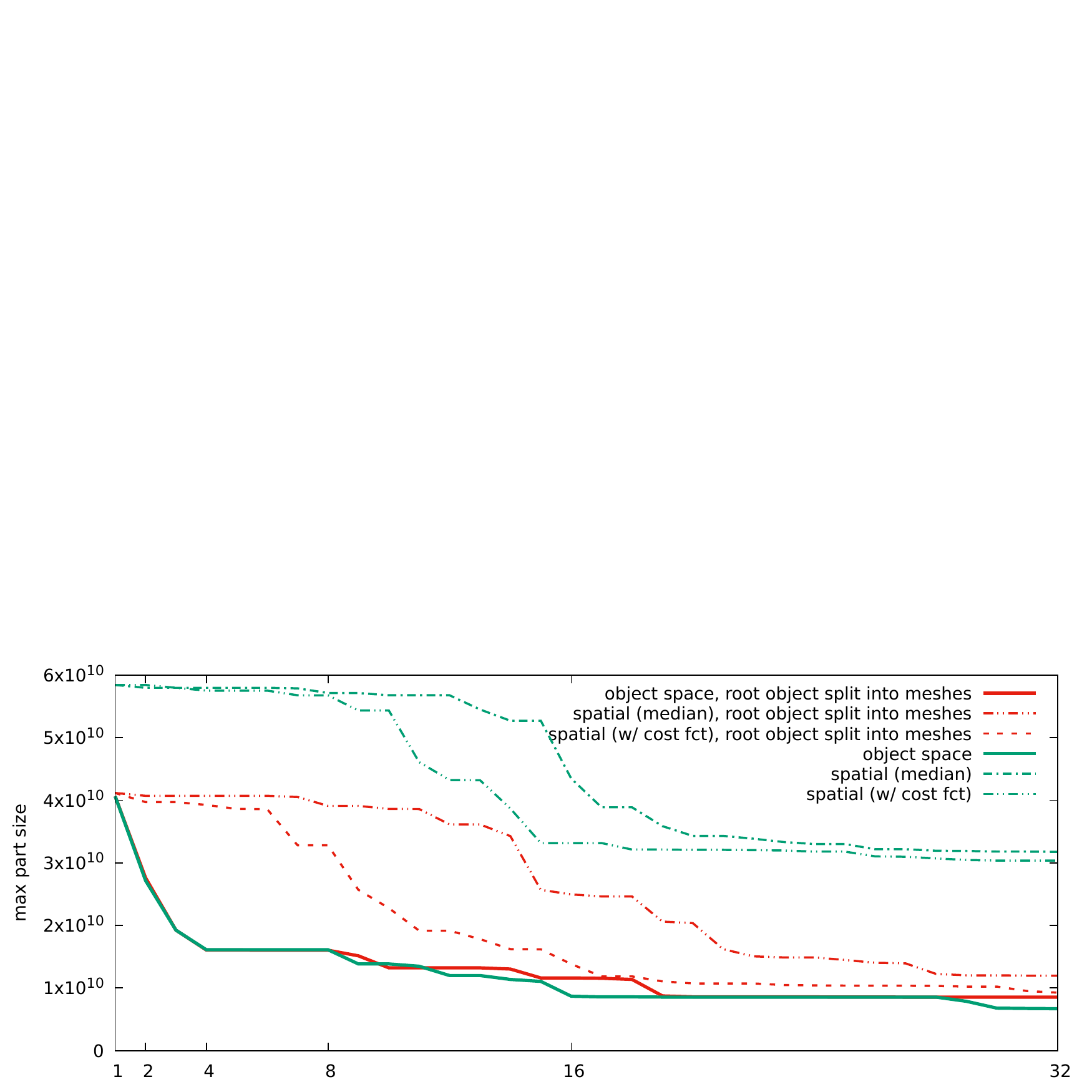}}
  \\[-1em]
  \caption{\label{fig:scale-max-part-size} Estimated memory usage of
    the largest part when splitting \island 
    into any number from 1 to 32 parts, for different
    partitioning strategies; 
    once with all non-instanced geometry together in
    one instance (green), and once with non-instanced objects broken into meshes
     (red).
    \vspace*{-4ex}
  }
\end{figure}

The result of this experiment are plotted in
Figure~\ref{fig:scale-max-part-size}: for purely spatial partitioning
the first few splitting planes hardly manage to reduce max part size
at all, even when we use multiple candidate planes; and even after
splitting into 16 parts has barely been reduced by $2\times$.  When
pre-splitting the single-instance root object into its
constituent meshes as proposed by Zellmann et
al.~\shortcite{zellmann} the situation markedly improves (red lines),
but even then it takes a lot of partitions to significantly reduce
model size. For object partitioning max model size drops
rapidly, and even with only 4 parts is almost as good as 
spatial partition is with 16; it also and eventually reaches a minimum that spatial
partitioning cannot reach.

For our renderer that means that the object-space and hybrid
partitionings will easily fit on the 8 GPUs we got for this
experiment, and even for \island only require 25~GBs out of the 48~GBs
available.  For the spatial partitions even with pre-splitting some of
the GPUs will temporarily exceed their memory during acceleration
structure build, which we currently allow by using CUDA managed memory
to temporarily allow paging out of data while the scene is built.

\subsection{Ray-Bandwidth Per Frame}

The promise of our motivational example from Section~\ref{sec:idea}
was that our techniques would not only help with how effectively a
scene could be partitioned, but would ultimately even reduce ray
bandwidth.  To evaluate this we instrumented our MPI code to track,
across all nodes and ray exchanges, how many such ray forwards were
required to render a given frame. In Table~\ref{tab:ray-bandwidth} we
report these numbers for the configurations and 
viewpoints shown in Figures~\ref{fig:proxies-per-model}
and~\ref{fig:views-per-model}: Na\"ive object space partitions incurs
ray bandwidth several times higher than spatial techniques, but the
introduction of proxies can reduce that significantly. Our currently
best partitioner---which by default is allowed to replicate up to
5\% of the input geometries---can do even better, and eventually
requires $2-35\times$ \emph{less} ray bandwidth than our best
spatial techniques.

In practice, this result is important, for two reasons: First, for a
system limited by how many rays can be sent across the network  any reduction in ray bandwidth directly
translates into higher frame rate.  For the default view of \island,
using the same hardware resources our best object-space
technique renders at 7.9 frames per second vs only 3~fps for
spatial; for the view that captures the whole model, these 
are 11.1~fps vs 3.7~fps, respectively. When using more than one sample
per pixel these speedups are even higher, as the bandwidth required for adding
the final frames becomes  less relevant.

Second, we observe that since our object space techniques require
less GPU memory per rank (25~GB vs 48+GB, see
Section~\ref{sec:max-part-size}) we could actually have used fewer
ranks for these, likely achieving yet higher performance with less resources.
For example, for \landscape frame rate for our object techniques goes from
6.2~fps to 7.1~fps when going from 4 workers to 2.

\def\rel#1{\relsize{-1}{\color{red}$(#1\times)$}}
\def\m#1{{#1\relsize{-1}{M}}}
\def\best#1{{\color{OliveGreen}\m{#1}}}
\begin{table}[ht]
  \vspace{-1em}
  \centering{\relsize{-1}{
      \setlength{\tabcolsep}{2pt}
\begin{tabular}{l|cccc|cccc|ccc}
  &
  \multicolumn{4}{c|}{spatial only} & \multicolumn{4}{c|}{object} &
  \multicolumn{3}{c}{hybrid} \\
  & \multicolumn{2}{c}{sp.median} & \multicolumn{2}{c|}{cost fct}
  & \multicolumn{2}{c}{naive} & \multicolumn{2}{c|}{proxies}
  & \multicolumn{2}{c}{bvh-like} & best \\
  \hline
  \hline
  &\multicolumn{5}{c}{PBRT \landscape}\\
  \hline
  path      & \m{1.9} & \rel{4.4} & \m{1.6} & \rel{3.6} & \m{6.7} & \rel{15} &  \m{16} & \rel{36} & \m{4.3} & \rel{9.7} & \best{.45} \\
  view-3    & \m{2.4} & \rel{5.2} & \m{2.2} & \rel{4.8} & \m{5.6} & \rel{12} &  \m{33} & \rel{72} & \m{5.0} & \rel{11}  & \best{.46} \\
  top       & \m{.89} & \rel{7.6} & \m{1.5} & \rel{13}  & \m{1.7} & \rel{14} & \m{7.3} & \rel{62} & \m{1.5} & \rel{13}  & \best{.12} \\
  \hline
  \hline
  &\multicolumn{5}{c}{Moana \island}\\
  \hline
  default  &  \m{5.0} & \rel{3.8} & \m{8.0} & \rel{6.2} & \m{19} & \rel{15}  &  \m{7.4} & \rel{5.7} & \m{9.7} & \rel{7.5} & \best{1.3}  \\
  beach    &  \m{4.7} & \rel{1.7} & \m{7.3} & \rel{2.7} & \m{24} & \rel{8.7} &   \m{14} & \rel{5.0} &  \m{20} & \rel{7.4} & \best{2.7} \\
  top      &  \m{8.6} & \rel{43}  & \m{7.3} & \rel{36}  & \m{12} & \rel{59}  &  \m{6.9} & \rel{35}  & \m{6.1} & \rel{31}  & \best{.2}  \\
\end{tabular}}}
  \caption{\label{tab:ray-bandwidth}
    Rays forwarded across all nodes and bounces, for one path per
  pixel at $2560\times 1080$, using the given view and partitioning
  method (see Figures~\ref{fig:proxies-per-model}
  and~\ref{fig:views-per-model} for reference).
  \vspace*{-3em}}
\end{table}

\subsection{Application to Non-Instanced Models}

To ascertain that our method is not limited to heavily instanced
models we also developed a separate content pipeline that can handle
large non-instanced models in an out of core fashion, creating the
same spatial partitioning that existing data parallel renderers (e.g.,
in sci-vis) would produce. This works just fine; however, a full discussion
of this is beyond the scope of this paper.

\section{Summary and Discussion}

In this paper, we have proposed a new approach to data parallel
rendering that is more general than spatial partitioning; while
simultaneously leading to better results in all of memory use,
bandwidth, and performance. The downside of this generality is that it
does not automatically define what the ``best'' partitioning might be,
nor even a good heuristic for measuring the quality of a given
partition, or for predicting the ray bandwidth required by one. 
Fully understanding this will require more research.

Our prototype renderer is, to our knowledge, the first to successfully
demonstrate interactive yet data parallel path tracing on models of
that kind that uses ray forwarding (even for spatial partitioning); however, real production
renderers are more complex than ours, and moving these techniques into
practice will require significant engineering. This is also true once
we start using hardware with higher network bandwidth; this will
require significant systems engineering to overlap communication with
rendering, etc. We also can not yet handle techniques like
bi-directional path tracing, photon mapping, or multiple importance
sampling; some of this will require more work.
Finally, it would be interesting to back-port some of our techniques
into sci-vis, too; e.g., by using proxies on
top of natively distributed sci-vis content.

\section{Conclusion}

The methods described in this paper allow for a more general
description of how models can be represented in a data-parallel
context, and when integrated into a non-trivial path tracer are not
only more general, but can also achieve---simultaneously---lower
memory use, lower ray bandwidth, and higher performance.
Integrating these techniques into actual products will certainly
require more research and more engineering work; however, we do
believe that these techniques will significantly influence how future
data parallel path tracers will be built.

\bibliographystyle{ACM-Reference-Format} \bibliography{main}

\end{document}